\documentclass[letterpaper]{article}
\usepackage[margin=1in]{geometry}
\usepackage{times}

\usepackage{hyperref}       
\usepackage{url}            
\usepackage{booktabs}       
\usepackage{nicefrac}       
\usepackage{microtype}      

\usepackage{graphicx} 
\usepackage{subfigure} 
\usepackage{natbib}
\usepackage{amsmath}
\usepackage{bm}
\usepackage{amsfonts}
\usepackage{amssymb}

\usepackage{algpseudocode,algorithm,algorithmicx}
\algdef{SE}[DOWHILE]{Do}{doWhile}{\algorithmicdo}[1]{\algorithmicwhile\ #1}%

\title{Adversarial Vulnerability Bounds for Gaussian Process Classification}
\author{\Large{\textbf{Michael Thomas Smith,\textsuperscript{\rm 1}\thanks{Corresponding author}\;\; Kathrin Grosse}},\textsuperscript{\rm 2}  \\\Large{\textbf{Michael Backes,\textsuperscript{\rm 2} Mauricio A Alvarez Lopez\textsuperscript{\rm 1}}}\\\textsuperscript{\rm 1}Department of Computer Science, University of Sheffield\\m.t.smith@sheffield.ac.uk \& mauricio.alvarez@sheffield.ac.uk\\\textsuperscript{\rm 2}
CISPA Helmholtz Center for Information Security, Saarland Informatics Campus\\kathrin.grosse@cispa.saarland \& backes@cispa.saarland}

\begin{document}

\maketitle

\begin{abstract}
Machine learning (ML) classification is increasingly used in safety-critical systems. Protecting ML classifiers from adversarial examples is crucial. We propose that the main threat is that of an attacker perturbing a \emph{confidently classified} input to produce a \emph{confident misclassification}. To protect against this we devise an \emph{adversarial bound} (AB) for a Gaussian process classifier, that holds for the entire input domain, bounding the potential for any future adversarial method to cause such misclassification. This is a formal guarantee of robustness, not just an empirically derived result. We investigate how to configure the classifier to maximise the bound, including the use of a sparse approximation, leading to the method producing a practical, useful and provably robust classifier, which we test using a variety of datasets.
\end{abstract}

\section{Introduction}
A machine learning (ML) classifier is given some training points $\mathbf{X}$ with labels $\bm{y}$, and is then asked to classify a new test point at $\bm{x}_*$. Modern methods now exist that can successfully classify a test point in high dimensional, highly structured datasets \citep[e.g.][]{NIPS20124824}. It has been found however that these successful classification results are susceptible to small perturbations in the location of the test point \citep{BiggioCMNSLGR13,szegedy2013intriguing}. By carefully crafting the perturbations an attacker can cause an ML classifier to misclassify even with only a tiny alteration to the original data point. 

There is currently an absence of proven bounds on adversarial robustness that provide guarantees across the whole input domain. We find we can acheive this however, using Gaussian process (GP) classification. Its two key benefits are, first, GPs allow one to specify priors, providing an equivalence to the smoothness mentioned by \citet{ross2018improving}. Second, GPs offer robust uncertainty quantification, allowing us to determine the confidence the algorithm has in a given prediction. These two features allow us to bound the effect a perturbation can have on a prediction.

Typically, to illustrate an Adversarial Example (AE), one might take some point which humans can easily classify, then perturb it (while minimising a chosen norm on the perturbation) to cause the ML algorithm to misclassify, crossing the decision boundary. 
We instead follow recent approaches of high-confidence adversarial examples \citep{carlini2017towards,grosse2018limitations}, with the additional constraint that the initial, starting point is also confidently classified. To motivate, if our self-driving car is only 55\% confident the traffic light is green, we would expect it to stop. We are more interested in AEs in which the classification is moved from, say, 99\% confident of a red light to 99\% confident of a green.

This work provides a guarantee that is considerably more universal than previous work, as it holds for the whole domain and lower bounds the scale of a perturbation required to change \emph{any} confidently classified, as yet unknown, future test point to the opposite class. This guarantee we refer to as the adversarial bound (AB). This AB method targets perturbations bounded by the $L_0$ norm as a common constraint on typical real-life attacks, where the adversary is likely to only have access to a small portion of the domain, e.g. stickers on road-signs. Other papers
chose norms to reflect human perception, but we feel the choice should also consider the capacity of the attacker. We propose that the $L_0$ norm reflects the most likely real-world attack. For example, \citet{sitawarin2018rogue} offer real-life image examples of road signs that effectively minimise this norm.
\citet{su2019one} show single pixel perturbations are sufficient to cause a misclassification with a typical DNN.

To summarise, we provide a formal guarantee of robustness against AEs and by so doing develop a robust classifier. In particular we use the concept of confident misclassification AEs and propose that these are what we should protect against.
Importantly the paper provides a true mathematical lower bound on the scale of perturbation required, not an empirically derived result. We use Gaussian process classification (GPC), a powerful and widely adopted classifier, as the basis for our analysis and demonstrate the AB method provides meaningful bounds for a variety of datasets, and how adjusting various parameters allows one to trade off accuracy and computation for robustness.

\subsection{Gaussian Process Classification}

A GP is a stochastic process defined such that any finite subset of output variables will have a multivariate normal distribution. The covariance between pairs of variables is defined by the choice of kernel function, $k(\mathbf{x},\mathbf{x}')$. 
One can condition on some variables to make a closed-form prediction of the posterior for the remaining. 
To be specific; we are given a set of training input-output pairs, $\mathbf{X} \in \mathcal{R}^{N \times D}$ and $\mathbf{y} \in \mathcal{R}^{N}$ respectively. We wish to estimate the values of the latent function $f$ at a test point $\mathbf{x}_*$. 
%
%
%
The conditional distribution of $f(\mathbf{x}_*) | \mathbf{y}$ can be expressed analytically as a normal distribution with mean and covariance, $\bar{f}(\mathbf{x}_*) = \mathbf{k}_*^\top (\mathbf{K} + \sigma^2\mathbf{I})^{-1} \mathbf{y}$ and $
\mathbb{V}[f(\mathbf{x}_*)] = \mathbf{k}_{**} - \mathbf{k}_*^\top (\mathbf{K} + \sigma^2\mathbf{I})^{-1} \mathbf{k}_*.$ Where $\mathbf{k}_{**}$ is the kernel variance at the test point. $\mathbf{k}_{*}$ and $\mathbf{K}$ are the covariances between test-training points and within training points respectively. The posterior mean can be written as the sum of weighted kernels due to the representer theorem 
\begin{equation}
\bar{f}(\bm{x}_*) = \sum_{i=1}^N \alpha_i k(\bm{x}_i,\bm{x}_*),
\label{representer}
\end{equation}
where $\bm{\alpha} = \bm{K}^{-1} \bm{y}$. 

To extend to classification, the likelihood will no longer be normal, but instead the probability of a point being of a given class (given the latent function).
For (binary) classification we still use a real-valued latent function, $\mathbf{f}$, but squash this through the logistic link function to give us the class probabilities, $\pi(\mathbf{x}_*) = \int \sigma(f_*)p(f_*| \mathbf{X}, \mathbf{y}, \mathbf{x}_*) df_*.$ 
The posterior prediction of the latent function is computed by combining the latent function and the likelihood and marginalising the latent function; 
$p(f_*| \mathbf{X}, \mathbf{y}, \mathbf{x}_*) = \int p(f_*| \mathbf{X}, \mathbf{x}_*, \mathbf{f}) p(\mathbf{f}| \mathbf{X}, \mathbf{y}) d\mathbf{f}$.
Neither of the above integrals can be solved analytically. 
In this paper we use the Laplace approximation to the posterior. Specifically we place a Gaussian $\mathcal{N}(\mathbf{f}|\hat{\mathbf{f}},A^{-1})$ on the mode $\hat{\mathbf{f}}$, of the posterior $p(\mathbf{f}|\mathbf{x},\mathbf{y})$ with a covariance $A^{-1}$ that matches the posterior's second order curvature at that point. Finding the mode and covariance is mildly involved and is described in \citet[p42-43]{williams2006gaussian}. The approximate \emph{mean} of the latent function is then computed as for normal GP regression, but using the mode. To compute the variance requires an additional integration step
but the AB-algorithm, for simplicity, ignores this source of uncertainty, and only considers the mean function. The additional uncertainty will only increase the perturbation required, so our algorithm still provides a lower bound.


\subsection{Confident Misclassification}
\begin{figure}
  \centering
  \includegraphics[width=0.45\textwidth]{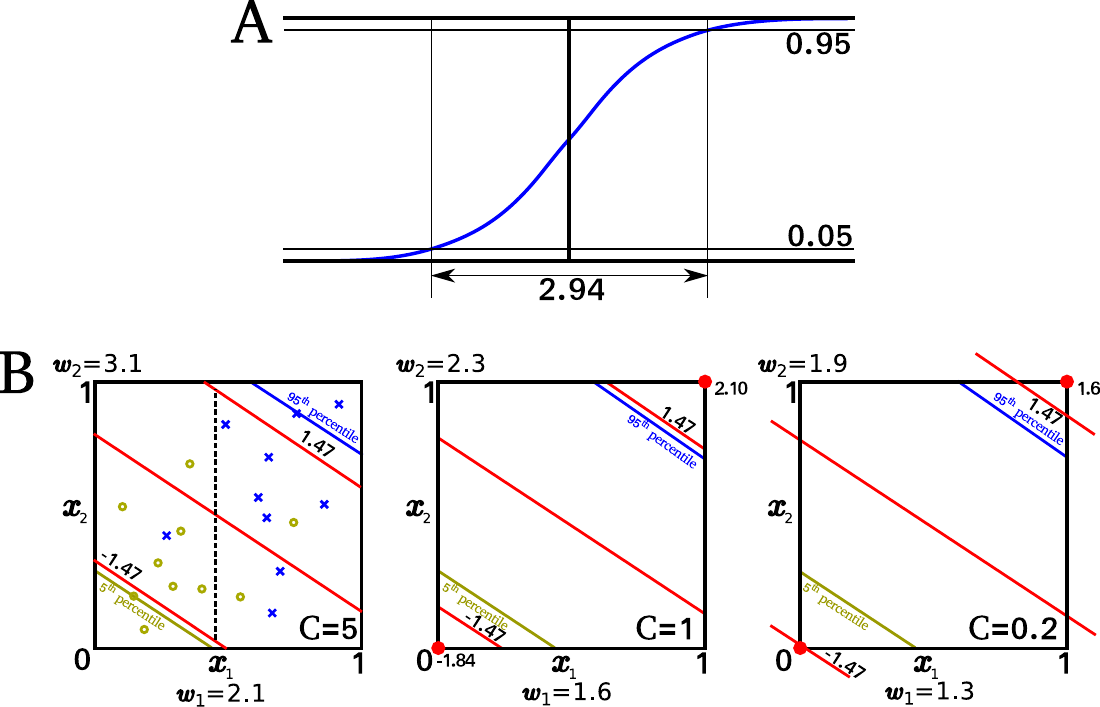}
\caption{2d example using LR. A) The logistic function, to move from $0.05$ to $0.95$ confidence in the prediction requires a change of at least $2.94$ in the latent function. B) The latent function $w_0 + x_1 w_1 + x_2 w_2$. 
Left, least regularisation: from $-1.47$ to $+1.47$ thresholds possibly by only changing just $x_2$ (as $w_2>2.94$). Middle \& right plots, greater regularisation, it becomes impossible to move between the two thresholds (i.e. when $w_1+w_2<2.94$) even changing all inputs! We suggest using the values of the 5th and 95th percentile training points (orange and blue lines) to assess the robustness of a classifier. $C$ is inverse $L_2$ regularisation.}
  \label{explanation}
\end{figure}

Rather than just produce a misclassification, we want to place a bound on the possibility of a \emph{confident misclassification}, i.e. moving from a confidently classified point of one class to a confidently classified point of another.
We demonstrate this using a simple example. We train a Logistic Regression (LR) classifier on a balanced two-class MNIST problem (100 digits, 3 vs 5, 8 $\times$ 8, with inputs normalised between 0 and 1) which gives 84.5\% accuracy on a test set, using $L_2$ regularisation, $C=1$. The 0.05 and 0.95 outputs are separated by 2.94 in the latent function (through the inverse logistic) so we could cumulatively sum the sorted weight vector. We find we need to alter at least three pixels to achieve at least a 2.94 change in the latent function (the largest two coefficients are 1.26 and 1.14). 
One can reduce the coefficients by simply increasing the regularisation, until the classifier can never reach the 0.95 threshold output (figure \ref{explanation}). Clearly this makes little sense. Instead we propose that one should use the value of the 5th and 95th percentile training points' values, and report how many inputs need perturbing to pass between \emph{these} two thresholds. This normalises the target of the adversarial attack by correcting for the scaling caused by regularisation. 
Regularising can still help protect the classifier from adversarial examples even using this new way of framing the problem, as will be illustrated later.
%


\subsection{Related Work}
\label{related_work}
Much of the focus of the adversarial ML field has been on AEs with respect to deep neural network (DNN) classification. Overfitting, linearity and sharp decision boundaries associated with DNNs are hypothesise to be the cause of their susceptibility \citep{papernot2016distillation}. Attempts have been made to regularise the DNN's output. \citet{ross2018improving} regularised the input gradients as they hypothesised a model with `smooth input gradients with fewer extreme values' will be more robust. \citet{papernot2016distillation} used distillation
 but this was later found to be insufficient \cite{carlini2017towards}. The conclusions of \citet{carlini2017adversarial} are that adversarial perturbations are very challenging to detect or mitigate: attempts at mitigation quickly being compromised. 
To move beyond the `arms-race', researchers have begun looking at how to provide formal guarantees on the scale of the perturbations required to cause a misclassification \cite{wong2018provable,huang2017safety,madry2018towards,hein2017formal,carlini2017provably}. These approaches only guarantee a ball around each training point is protected from AEs. For example \citet{wong2018provable} propose a method for producing a bound on the scale of adversarial examples if using a deep (ReLU) classifier. 
\citet{hein2017formal} also provide a formal proof for a minimum perturbation $\delta$ which could create an AE around a particular training point, with a convex norm. 
Recently however, research has been conducted into the change that can be brought about due to perturbations \emph{anywhere} in the domain. This ensures that future, unseen test data has guarantees regarding the minimum perturbation required to cause a misclassification. \citet{peck2017lower} find a formal bound for DNNs in which one can start at any location, but produce bounds that are many orders of magnitude smaller than the true smallest perturbation. 
 \citet{cardelli2019robustness} 
find bounds on the probability of a nearby point to a test point having a significantly different prediction. 
We found that the above results did not provide practical bounds for use across the whole domain, or with an $L_0$ norm attack. The method described in this paper achieves both.
%

\section{Method}
\subsection{The Adversarial Bound (AB) Algorithm}
\label{alg}
\begin{figure}
  \begin{center}
  \includegraphics[width=0.45\textwidth,trim=0 15 0 15]{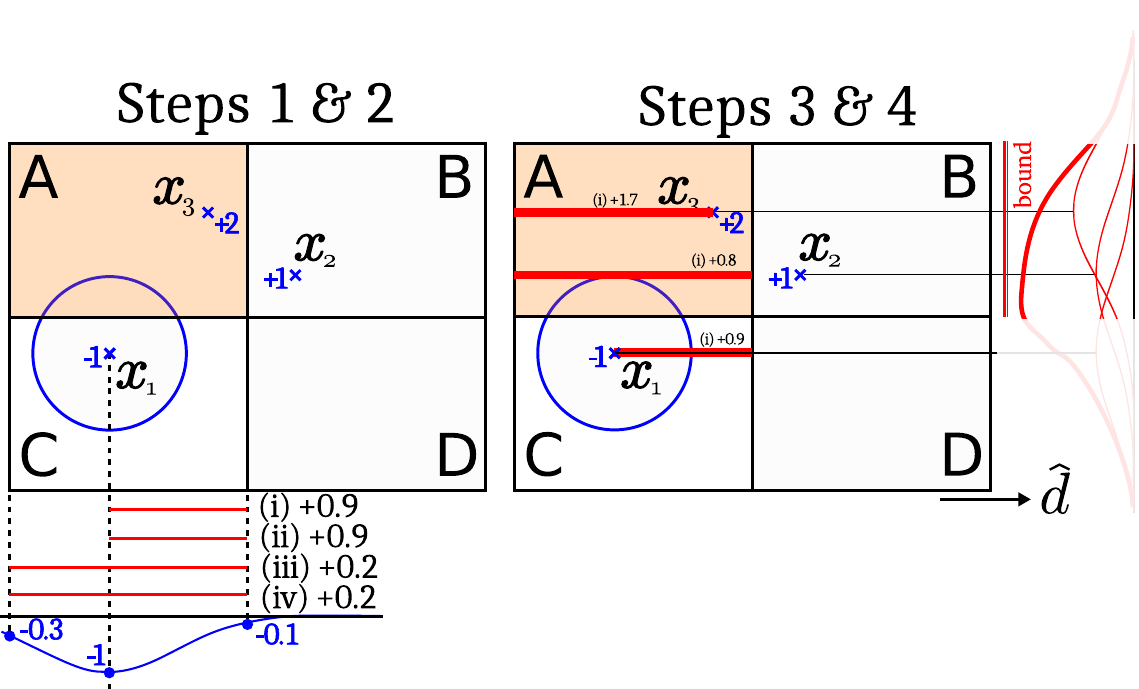}
  \end{center}
\caption{\emph{\textbf{Step 1}}, the 2d domain is sliced into four hypercubes. \emph{\textbf{Step 2}}, consider the negative training point $\bm{x}_1$ and hypercube A. Paths along dimension $\hat{d}$ (positive direction) through $x_1$ that maximise the increase in $\alpha_1 k(\bm{x}_1,\bm{x}_*)$ are indicated with the four possible constraints, (i)-(iv).
\emph{\textbf{Step 3}}, Combine bounds from hypercubes on path. If we the path that \emph{starts in A and ends in B}, we would add together (ii) from A and (iii) from B for each input. For paths starting and ending in A we consider just the values of (i) for each input, for that hypercube.
\emph{\textbf{Step 4}}, we find an upper bound on the maximum of this weighted sum.
}
  \label{algorithm1}
\end{figure}
To summarise the method, we consider perturbing each dimension, $\hat{d}$ of the $D$ dimensions, one at a time. We slice up the domain into smaller hypercubes, and consider all possible orderings of these hypercubes when travelling along a path in direction $\hat{d}$ (typically we just divide the domain into slices parallel to this direction). We wish to find the largest increase each hypercube can contribute to the posterior as part of a path. Each hypercube may be at the start, middle or end of the path, or may enclose the whole path. We sum these separate contributions which allows us to represent the possible increase as a weighted sum of exponentiated quadratic (EQ) kernels (of $d-1$ dimensions), assuming from here onwards that we are using an EQ kernel to define the GP prior. 
We step through this in detail below.
%
%

\renewcommand{\labelenumii}{\roman{enumii}}
\begin{enumerate}
\item Split the domain into (axis aligned) hypercubes.


\item Moving in a positive direction along the $\hat{d}$ dimension, inside each hypercube, for each term of the summation in equation \ref{representer}, we want to determine the largest possible increase the term can experience, centred on each training point (figure \ref{algorithm1}). Specifically, we want to find a upper bound on the increase in the posterior associated with a single training point, given four types of constraint, when moving the test point along dimension $\hat{d}$.

\begin{enumerate}
 \item starting and finishing anywhere inside the hypercube.
 \item starting inside and finishing on the upper edge.
 \item starting on the lower edge and finishing inside.
 \item crossing the whole hypercube.
\end{enumerate}
\item For the path segment we are trying to bound we select the appropriate (i)-(iv) values for each input, for each hypercube. For example if our path consists of four hypercubes in a row we would take (ii) from the first, (iv) from the second and third and finally (iii) from the fourth. Sum these contributions for each input. This gives the greatest contribution that a training point could cause to the posterior.
\item Construct a sum of EQs in $d-1$ dimensions, with each EQ centred on one training point and weighted by the above value. Find an upper bound on this sum of EQs within the domain of the path's hypercubes. We use the algorithm detailed in the supplementary material to find this bound.
\end{enumerate}
We repeat steps 3-4 for each possible sequence of hypercubes when travelling along dimension $\hat{d}$. The algorithm will need running again with the training point values reversed, to account for paths moving in the negative direction along $\hat{d}$.
To improve the above runtime we run a fast initial pass, with few slices. Then run the algorithm on those combinations with the greatest bound (i.e. capable of changing the latent function the most), using a higher slice count. We refer to this procedure as `enhancement'.

\subsection{Higher dimensions}
\label{higher_dimensions}
To find a lower bound on perturbing $n$ inputs, one can run the above algorithm, assigning $\hat{d}$ as each of the $D$ dimensions, before summing the largest $n$ of these. We found that for the datasets used, this straightforward method achieved reasonable results.
Alternatively, path sequences can be tested in the above framework that incorporate more than one dimension changing if the domain were sliced in multiple directions (e.g. one could test $A \rightarrow B \rightarrow C$ from figure \ref{algorithm1}). This will give a tighter bound, but at the cost of more computation, with ${D}\choose{n}$ combinations, which for small $n$ is approximately $\mathcal{O}(D^n)$ time. 
If $s$ is the number of slices we divide each dimension into. The overall time complexity is $\mathcal{O}((Ds^2)^n)$. 
In the algorithm above we are finding an upper bound on the effect each dimension can have on the latent function. By cumulatively summing these upper bounds, we find a lower bound on the number of dimensions that are required to change.

\subsection{Classification and the Sparse Approximation}
The algorithm above is for regression but we wish to consider GPC. For the Laplace approximation we find the mode and Hessian of the posterior. We can then use normal GP regression but with this alternative set of training values, $\bm{\hat{f}}$ in \citet{williams2006gaussian}[p44].

The bound becomes increasingly loose as the number of training points increases, however one can use sparse approximation methods \cite{snelson2006sparse} to mitigate this. Specifically, one replaces the original training data with a low-rank approximation using inducing inputs, a standard practice when using GPs with big data. First find suitable inducing point locations using gradient ascent to maximise the marginal log likelihood using the original $\bm{\hat{f}}$, then use the inducing inputs associated with the low-rank (e.g. deterministic training condition, DTC) approximation as the new training data for our classification. We then have $\bm{\alpha} = \sigma^{-2} \bm{\Sigma} \bm{K}_{uf} \bm{\hat{f}}$, 
where $\bm{\Sigma} = (\sigma^{-2} \bm{K}_{uf} \bm{K}_{uf}^\top + \bm{K}_{uu})^{-1}$ and $\bm{K}_{uu}$ is the covariance between inducing inputs, $\bm{K}_{uf}$ is the covariance between the inducing inputs and the original training points. 



\section{Results}
We consider several classification problems, and for each compare the GPC results to LR, considering both accuracy and robustness. 
We start with a simple 3d example then consider several MNIST examples and a non-separable synthetic dataset. We apply the method to three real classification datasets in which robustness against AEs is important for security, specifically credit-worthiness, spam-filtering and banknote-forgery. We investigate empirically the effect of the number of splits and the number of inducing inputs on the algorithm's robustness, accuracy and runtime, before finally generating AEs that reach our new confident misclassification threshold to demonstrate the scale of perturbation required.

\subsection{Three dimensional toy example}
\label{threed}

To demonstrate the method we consider a simple example using two hundred training points divided equally between two Gaussians ($\sigma = \frac{1}{4}$) located at $\Big[\frac{1}{4},\frac{1}{4},\frac{1}{4}\Big]$ and $\Big[\frac{3}{4},\frac{3}{4},\frac{3}{4}\Big]$ (kernel variance, $v=1$; lengthscale, $l=2$). 
We use the Laplace approximation and find the $\bm{\hat{f}}$ values for the training points. These values we then use in our normal GP regression. The classifier identifies 97 of the 100 test points correctly. We find that for this example, the training points that are in the bottom and top 5\% of the latent function are at -2.03 and +1.94, thus require 3.97 increase in the latent function to move from one to the other. We considered perturbations in 2 (of the 3) dimensions simultaneously, slicing the domain into a $4 \times 4 \times 4$ grid. Then on the 100 largest bounded solutions we ran a higher resolution ($16 \times 16$) two dimensional grid search to find more precise bounds.
Our AB algorithm provides an \emph{upper} bound on this possible change (using the above parameters) of 3.94. This is associated with a path that moves from the search cuboid at one corner of the domain to one at the opposite corner.
One might expect the start and end to lie at the centres of the two clusters but due to the longer lengthscale the latent function has its maxima and minima outside the input domain.
The value our algorithm found is just below the change required for a \emph{confident misclassification}. Hence we can guarantee that all three dimensions need to change to move from a confident classification of one class to a confident classification of the alternative class.

We also tested one million two-dimension perturbations. The largest change these perturbations caused was 3.49, less than the increase from the bottom 5th to top 5th percentile. This is a \emph{lower} bound on the change in the latent function.

\begin{figure}
  \centering
  \includegraphics[width=0.22\textwidth]{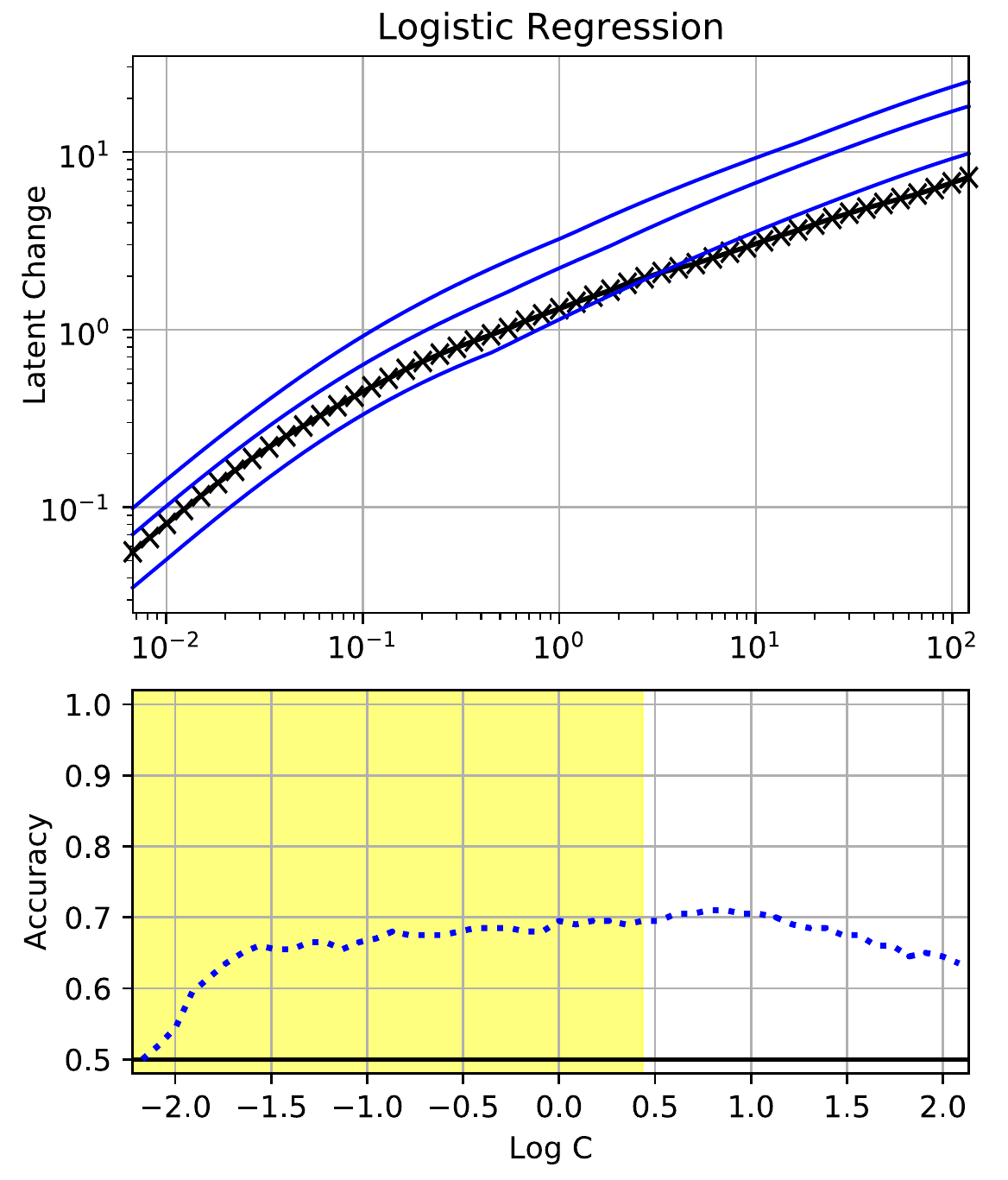}
  \includegraphics[width=0.22\textwidth]{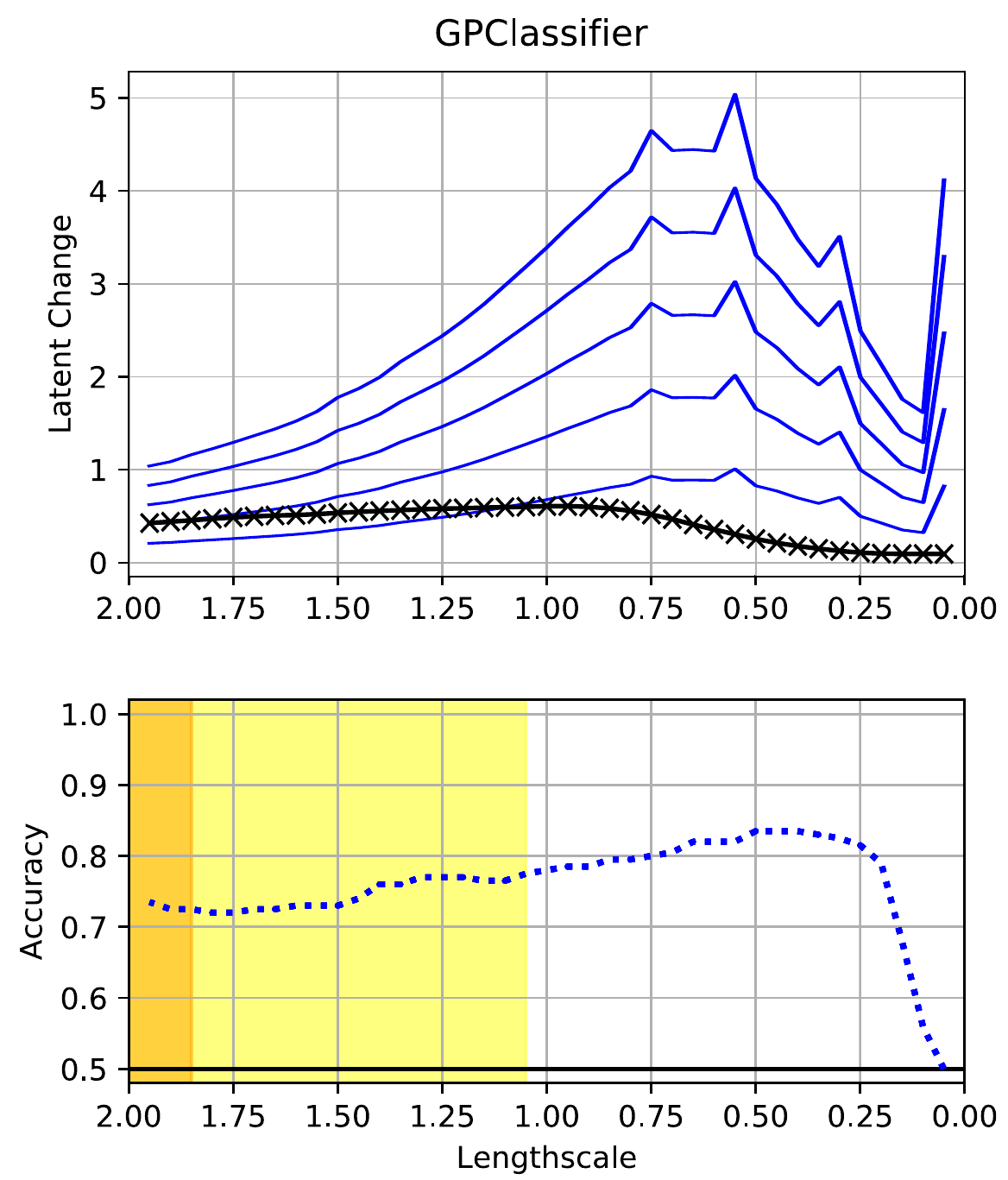}
\caption{MNIST 3v5 test using 100 training points. Logistic Regression (left) and GPC (right). Upper plots: The bounded change in the posterior induced by changing 1, 2 or 3 input points (blue lines) for given values of the regulariser or lengthscale. The black line indicates the `confident misclassification' threshold. Lower plot: how accuracy varies. The yellow/orange areas indicate regions in which two/three pixels need changing to cause a confident misclassification.}
  \label{mnist3v5_100data}
\end{figure}

\subsection{$8 \times 8$ MNIST (3 vs 5)}
Moving toward a real example we start with the relatively simple $8 \times 8$ MNIST, classifying $3$ vs $5$, using 100 training points. Only 43 pixels of the 64 are used by the classifier, as the remaining pixels were almost constant across training points. Figure \ref{mnist3v5_100data} shows that the GPC has a greater accuracy (over 80\%) than the LR algorithm (about 70\%). The GPC (with a lengthscale of two) achieves a bound of at-least 3 pixel-changes and a higher accuracy than the LR solution, which can only achieve a two pixel-change bound. Thus the GPC solution is more robust \emph{and} more accurate.
\subsection{$15 \times 15$ and full MNIST using inducing points}
\begin{figure}[tb]
  \centering
  \includegraphics[width=0.45\textwidth,trim=5 0 15 5,clip]{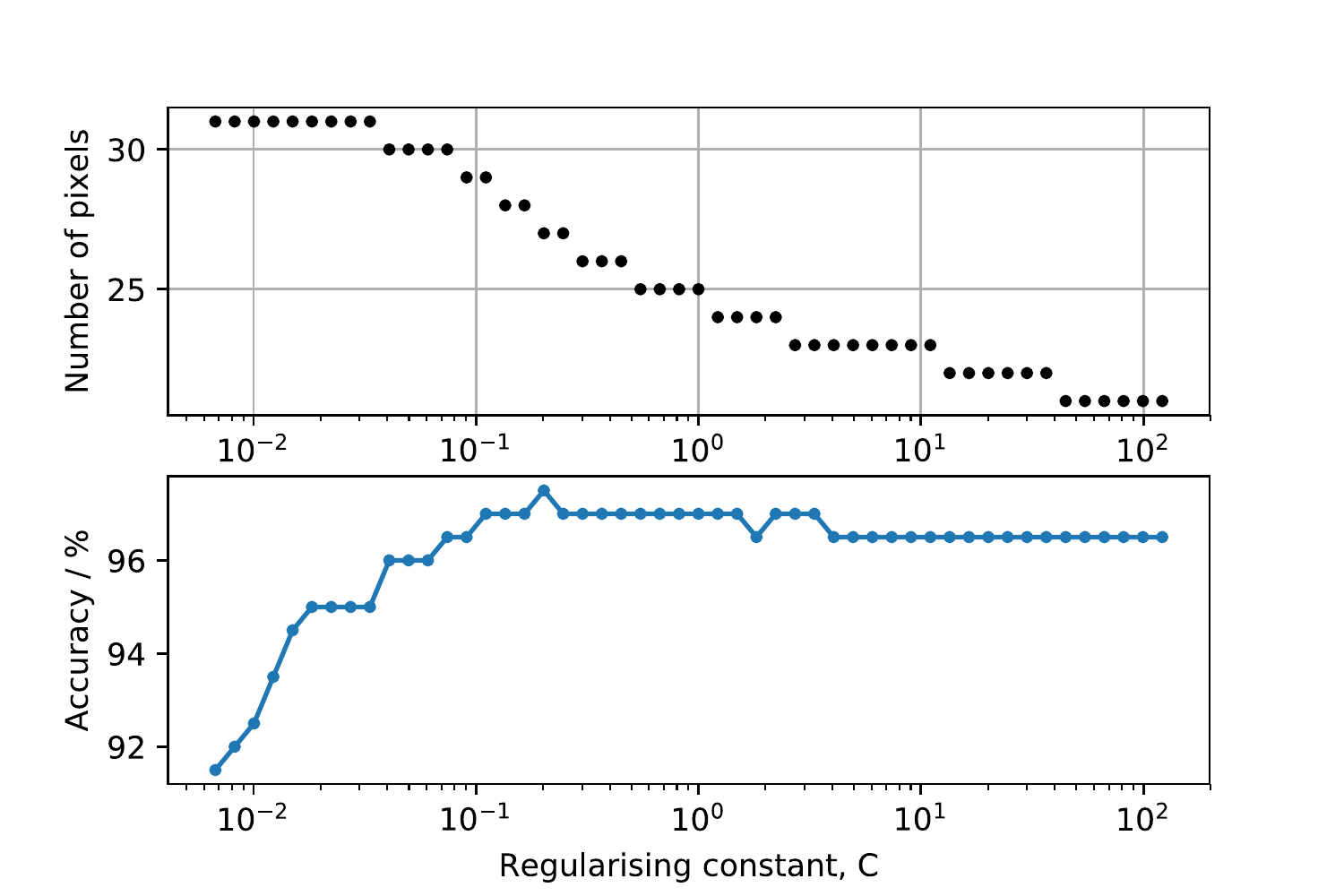}
  \includegraphics[width=0.45\textwidth,trim=5 0 15 15,clip]{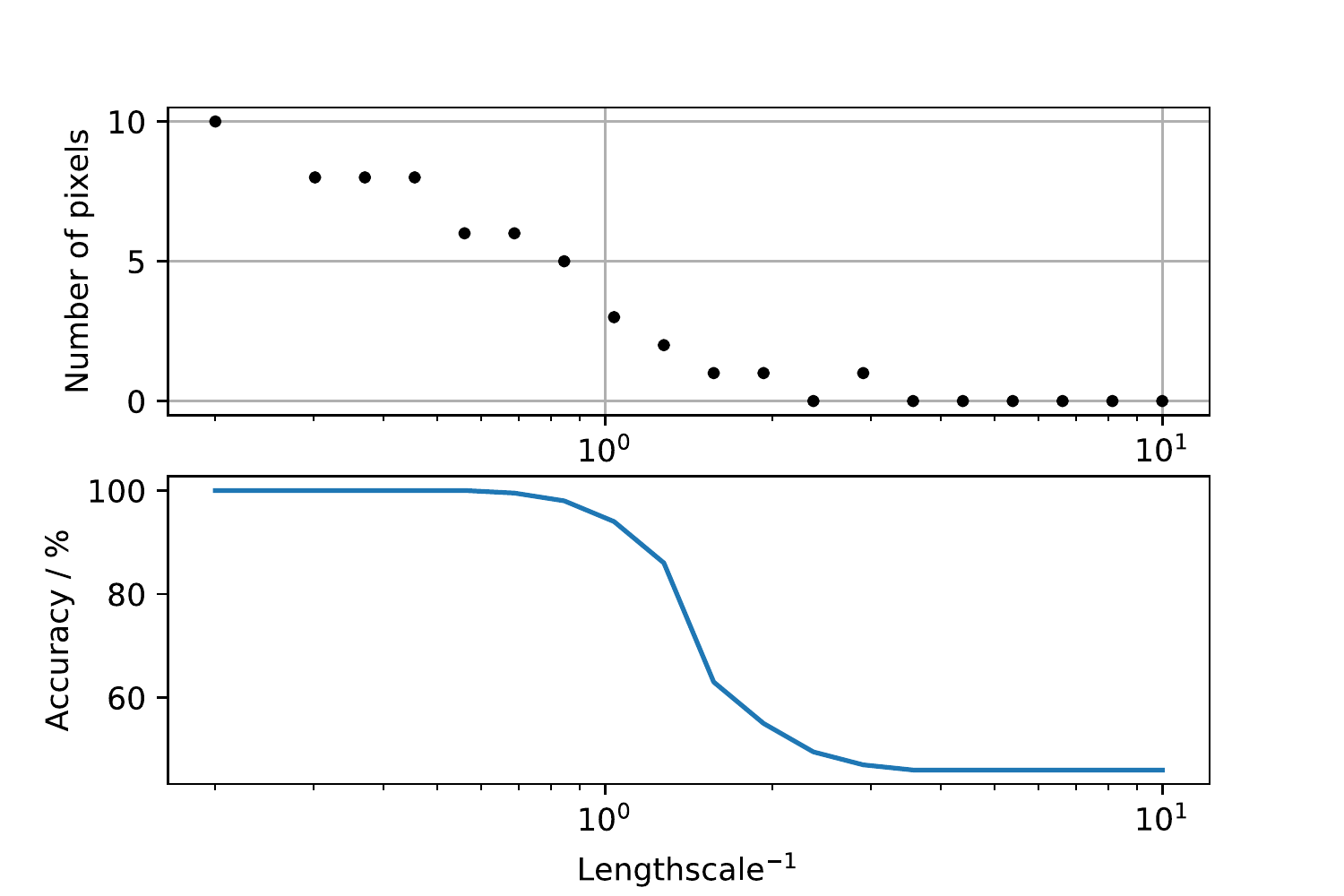}
\caption{Upper pair, LR; lower pair, GPC. $15 \times 15$ MNIST, 1 vs 0. Upper plots of each pair: the number of pixel changes to achieve a confident misclassification. Lower figures: Classifiction accuracy. Lengthscale inverted so increasing regularisation is to the left in both plots.}
  \label{mnist_1v0big_pix_change}
\end{figure}
\begin{figure}
  \centering
  \includegraphics[width=0.47\textwidth]{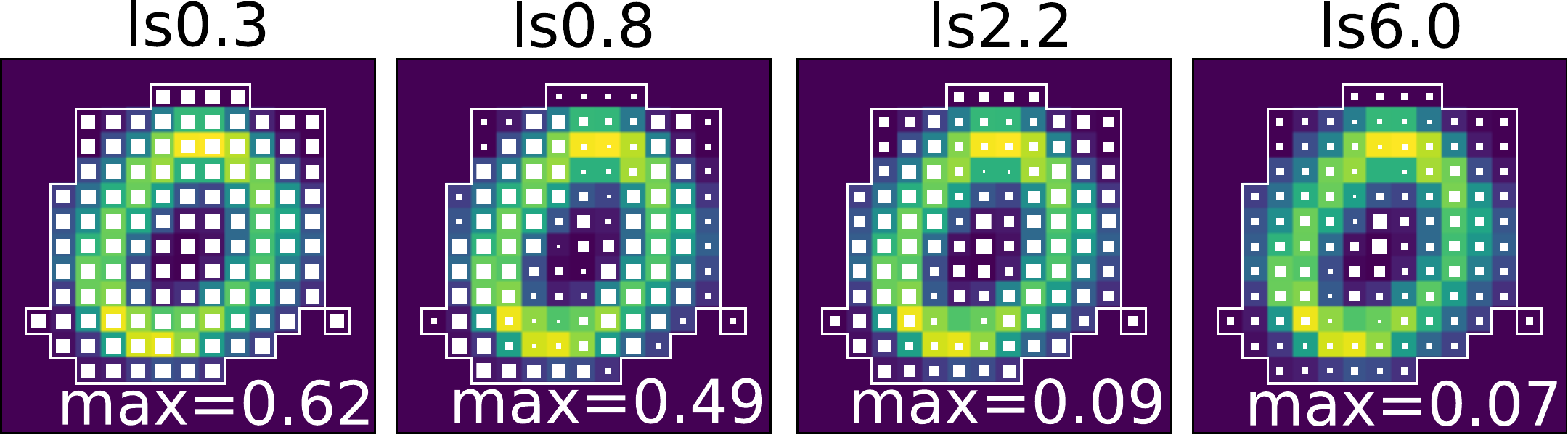}
\caption{MNIST 1 vs 0 with different lengthscales (4 inducing points \& 124 pixels used). Average images provide a visual reference. Squares show the (bounded) effect each pixel has on the posterior. Max is the largest square's value. Longer lengthscales lead to all pixels having a reduced effect on the posterior. Interesting structures are visible in the pixels identified as most perturbing.}
  \label{mnist_1v0big}
\end{figure}
We next apply the method with just four inducing inputs \& 100 training points on $15 \times 15$ MNIST (0 vs 1). Using just pixels that exceed 50 (of 255) leaves us with 124 of the 225 pixels as inputs. Probably due to the very separable classes, we found that both LR and GPC required many of the pixels to be changed to reach a confident misclassification threshold (figure \ref{mnist_1v0big_pix_change}). Figure \ref{mnist_1v0big} indicates the relative bound for each pixel, and how these are altered by changing lengthscale.

We finally applied the GPC AB method to the full resolution $28 \times 28$ MNIST images (475 of the 784 pixels were used as some pixels had little or no change between classes). 1000 training points (and four inducing points) were used.
The results are recorded in table \ref{fullmnist}. For some configurations dozens of pixels are required to change to achieve a confident misclassification.

\begin{table}
\centering
\begin{tabular}{p{1cm} p{1.2cm} p{2.1cm} p{2.5cm}}
Length-scale & Accuracy & Pixels required to change & distance between thresholds \\
\hline
2.154 & 0.905 & 15 & 12.525\\
10.000 & 1.000 & 76 & 4.916\\
21.544 & 1.000 & 44 & 3.534\\
46.416 & 0.990 & 58 & 2.541\\
\hline
\end{tabular}
\caption{AE algorithm results for $28 \times 28$ MNIST 0 vs 1. The accuracy is degraded by lengthscales that are too short or too long.}
\label{fullmnist}
\end{table}

\subsection{Non-separable synthetic data}
To demonstrate the AB algorithm on a dataset for which LR would fail, we generated an 8-dimensional linearly inseparable synthetic dataset of 50 training points placed in three Gaussian clusters ($\sigma=0.1$) within a unit hypercube, along its main diagonal. The LR classifier, as expected, fails to classify beyond chance.
The GPC however achieves 96\% accuracy ($l=0.7$, $v=0.3$, $\sigma^2=1$). 
With this configuration, the 5th and 95th percentile training points lie 0.321 apart. The AB method (using enhancement) found the upper bound for a single input perturbation was 0.220. 
Thus at least two inputs need perturbing to cause a confident misclassification.
A brute-force search found a change as large as 0.144 was possible with a single dimension change. Thus the true value lies between 0.144 and 0.220. 
See supplementary for more details of the trade off between accuracy and bounds for this example.


\subsection{Real world data: Credit, Spam and Banknotes}
\label{creditspambank}
\begin{figure}[t]
  \centering
  \includegraphics[width=0.22\textwidth]{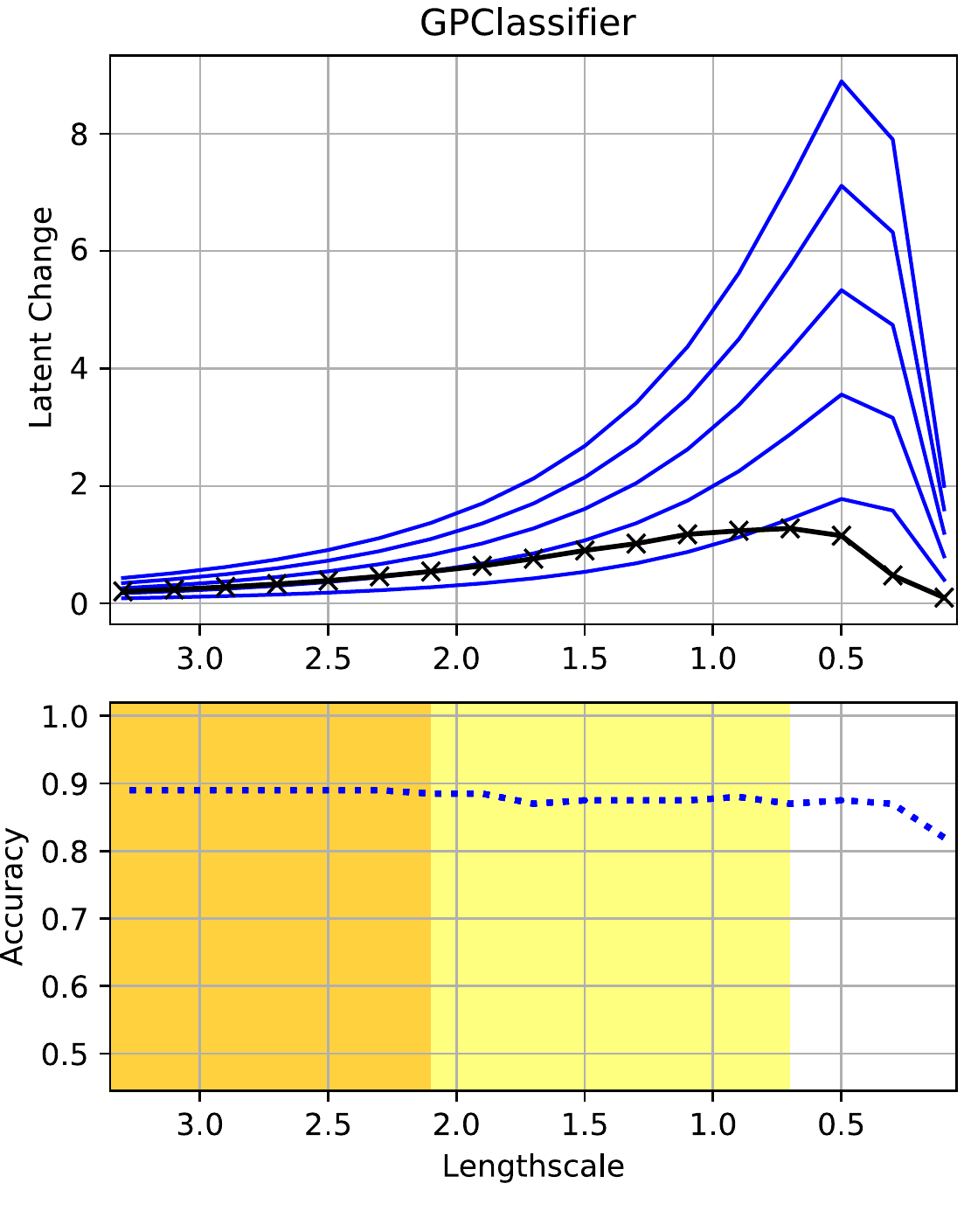}
  \includegraphics[width=0.22\textwidth]{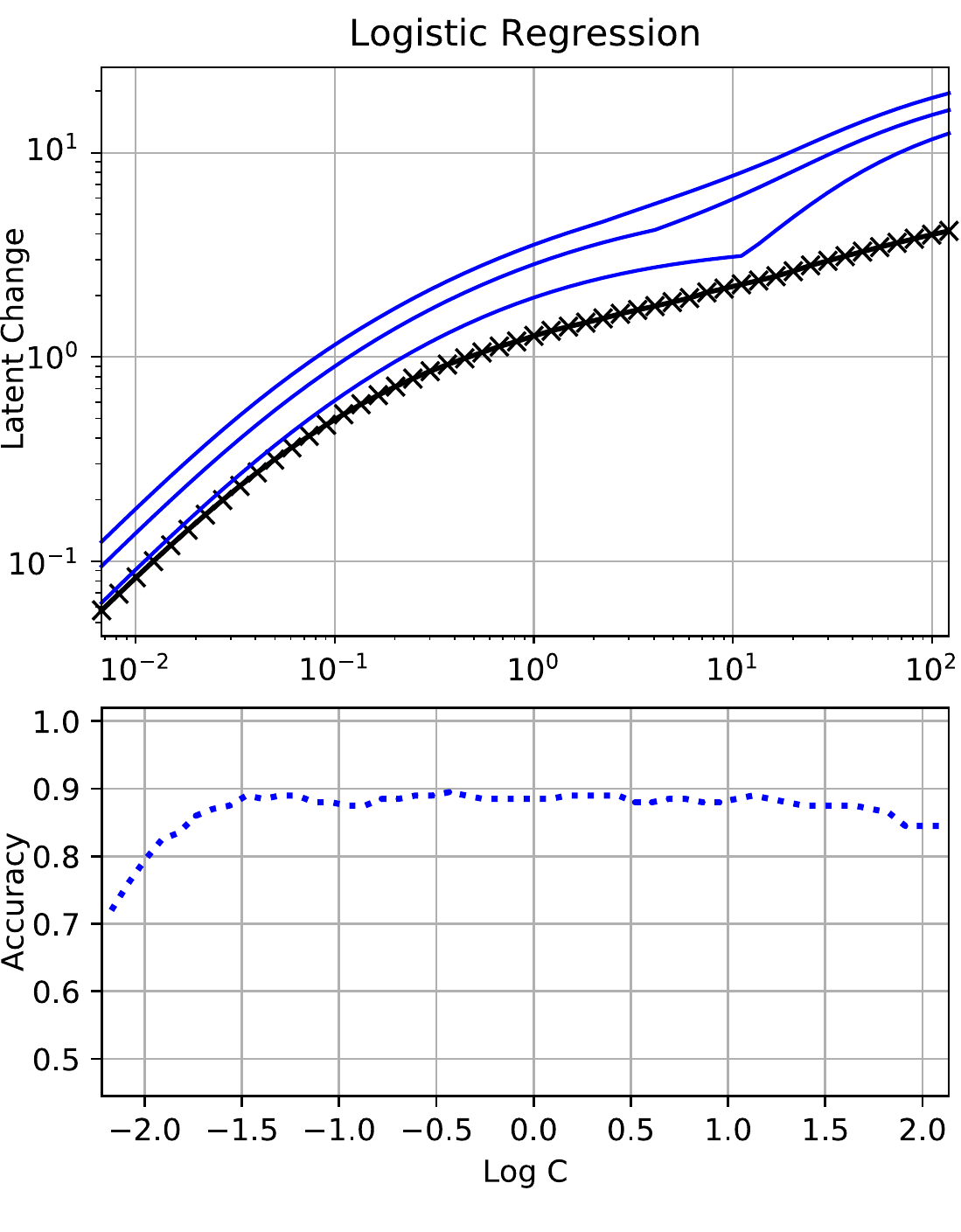}
  \includegraphics[width=0.22\textwidth]{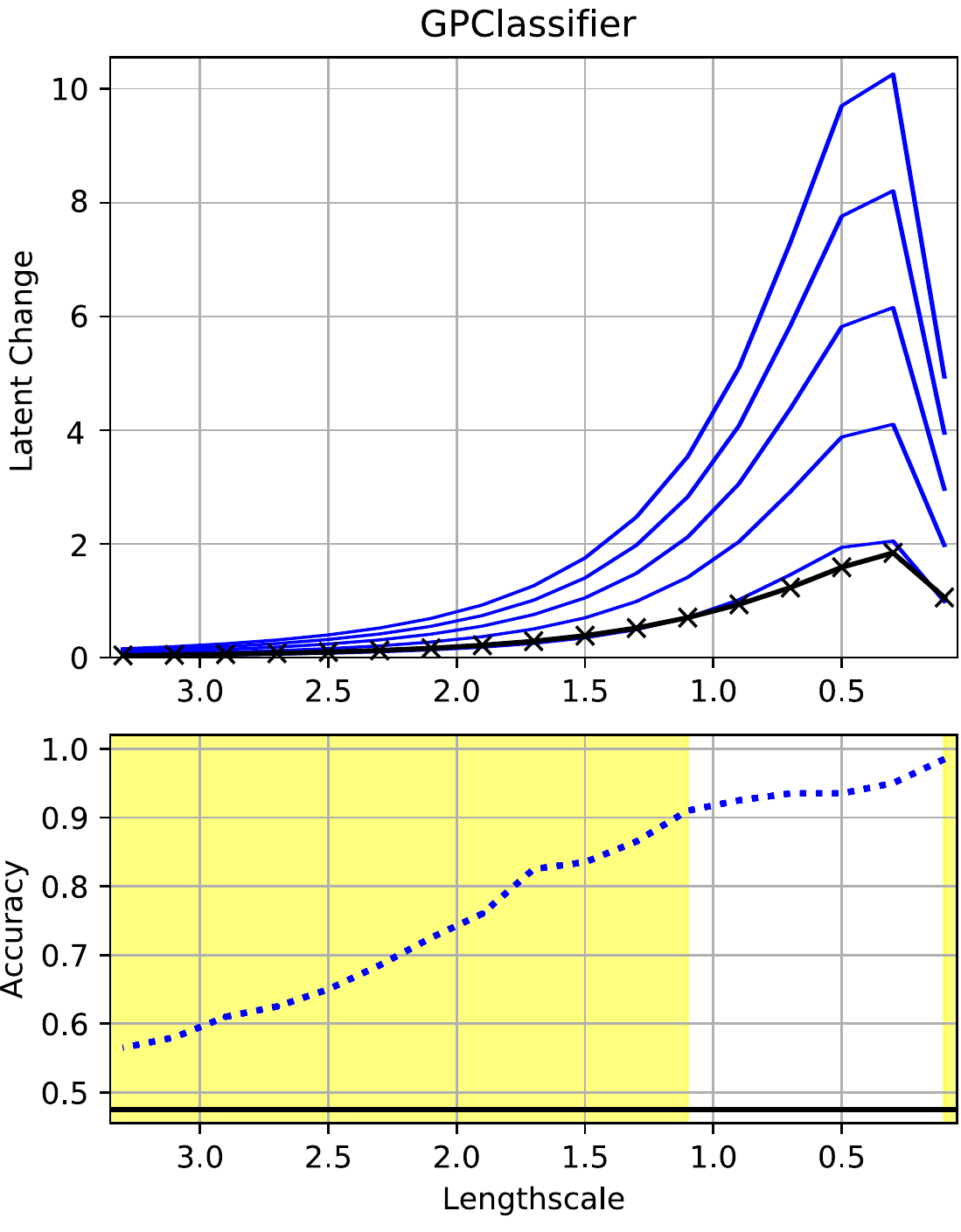}
  \includegraphics[width=0.22\textwidth]{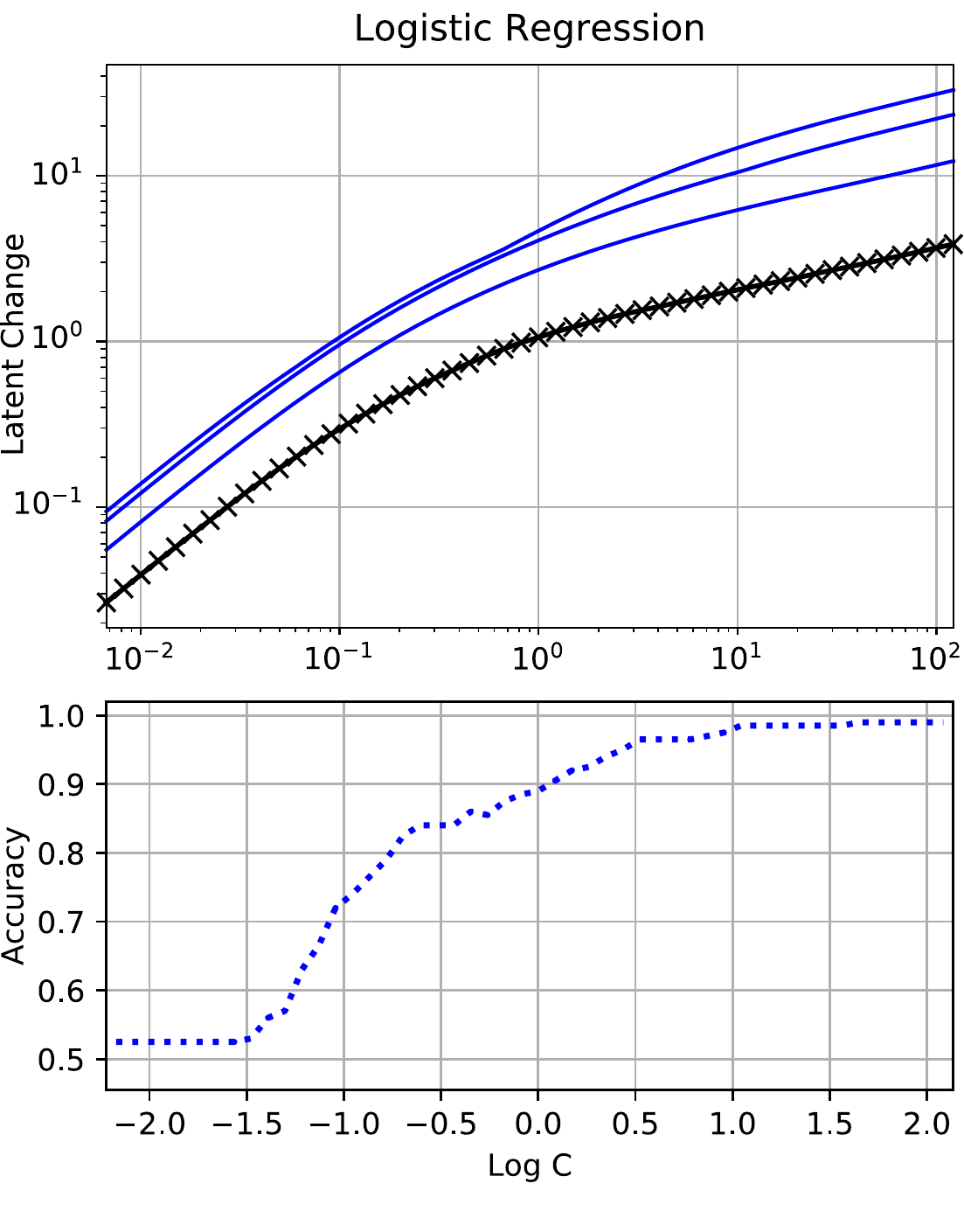}  
\caption{Credit (upper pairs) and bank dataset (lower) using GPC (left) and LR (right). Upper of each pair shows bounds on the impact on the latent function. Lower of each pair shows accuracy. Yellow/orange areas are where two/three pixels need changing to cause a confident misclassification. Black line indicates confident misclassification threshold.}
  \label{bankcredit}
\end{figure}
The `Australian Credit Approval' dataset contains 690 training points, each with 14 inputs consisting of binary, categorical and continuous data \cite{asuncion2007uci}. 
Upper plots in figure \ref{bankcredit} illustrate the analysis.
The fairly high accuracy for LR suggests the problem is mostly linearly separable, thus the long-lengthscale GP is able to maintain a good accuracy. It is unclear why it achieves more robustness compared to the linear classifier, the most likely explanation is that the two classes are somewhat compact.

We tested the algorithm on the spam dataset \cite{cranor1998spam} a 57 dimensional dataset. 
We found both the GPC and LR classifiers non-robust, i.e. both had a bound less than one input, providing no protection. Both methods achieved over 85\% accuracy (chance, 60\%). We suspect that some data are structured such that protection from AEs will be difficult to achieve. We also tested the algorithm on 100 points from the four dimensional UCI banknote authentication dataset. For the GPC, at least two of the four inputs needed changing for some lengthscales, while LR was not robust (Figure \ref{bankcredit}, lower plots). Note that at the shortest lengthscale the GPC is not only accurate but also more robust than at middle-lengthscales. 

\subsection{Effect of number of splits on bound and runtime}

The contribution of each training point is assumed to be the `worst-case' for a given hypercube. By introducing more hypercubes we tighten this bound.
\begin{figure}
  \centering
  \includegraphics[width=0.22\textwidth]{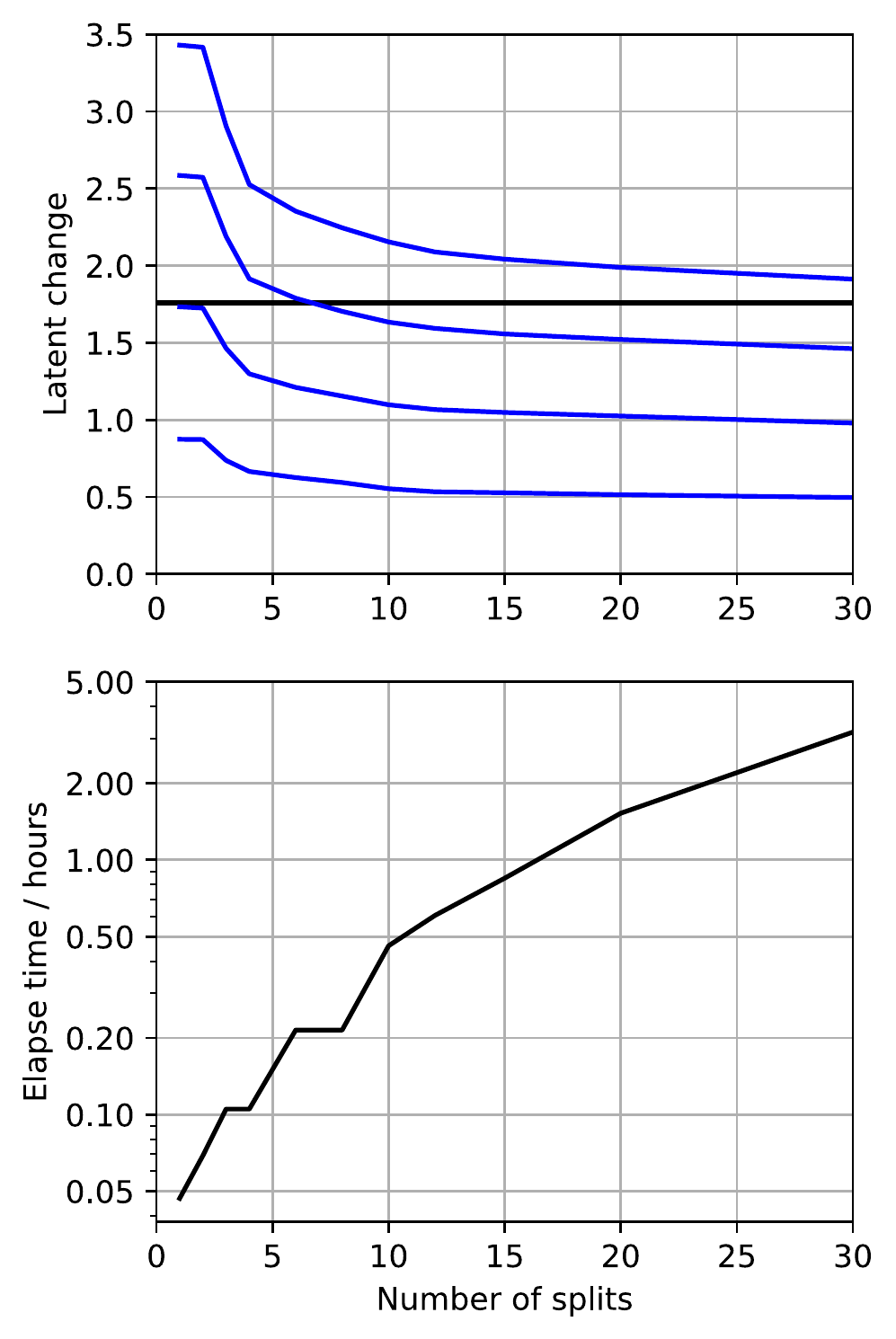}
  \includegraphics[width=0.22\textwidth]{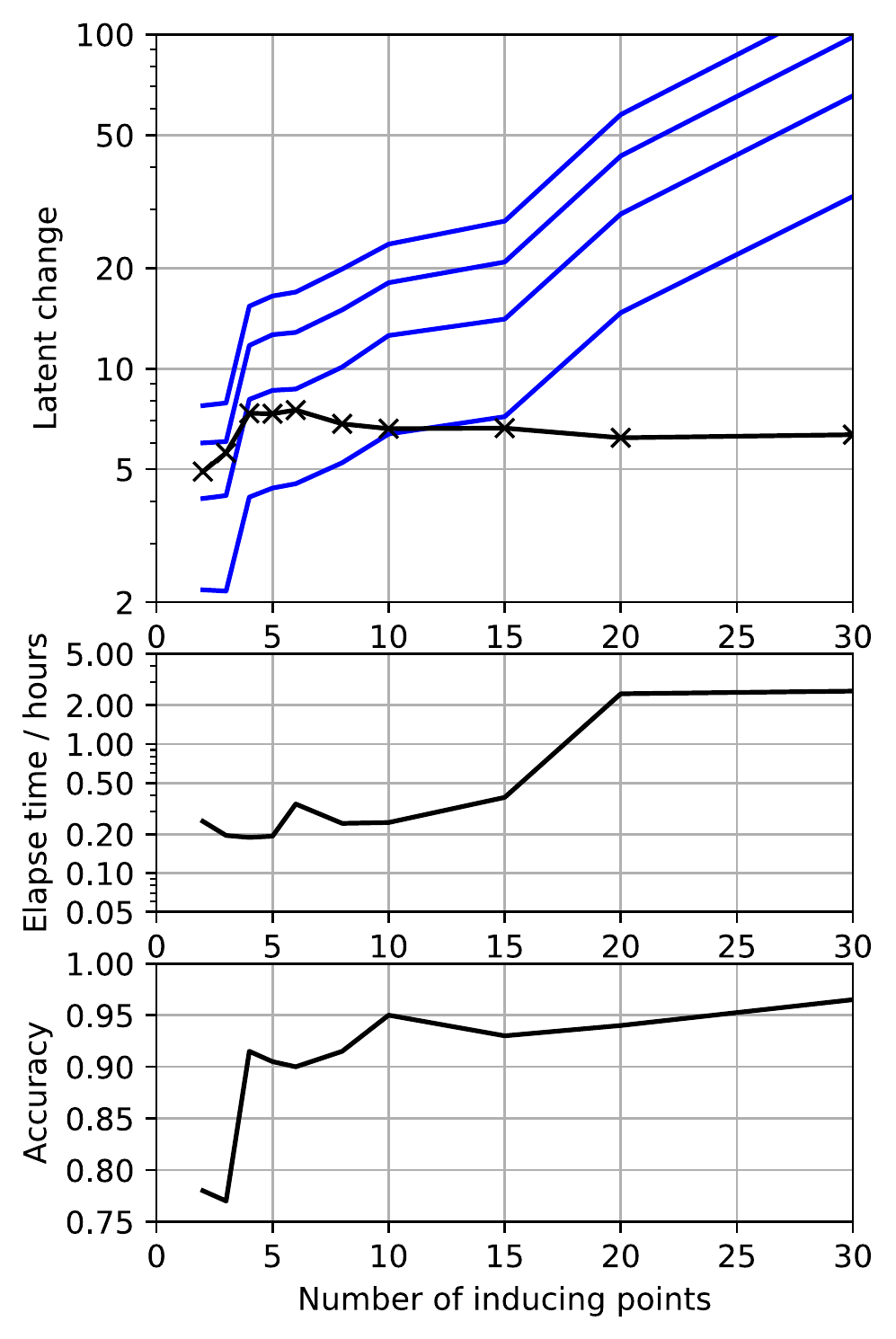}
\caption{Effect of (left) number of domain slices (right) number of inducing points. (upper) Cumulative effect of the first, two, three and fourth most significant input dimensions are indicated by blue lines. Black line indicates confident misclassification threshold (distance between the 5th and 95th percentile training points). For slicing experiment this was 1.758. See text for dataset and configuration details. Lower plots show elapse times and accuracy.}
  \label{sparseandsliceeffect}
\end{figure}
To test this effect empirically, we consider again the $8 \times 8$ MNIST (3 vs 5) data (100 training points, 200 test points, $l=4$, $v=1$, accuracy:68.5\%). The 5th and 95th percentiles lie 1.758 apart. Left plots in figure \ref{sparseandsliceeffect} demonstrate how the number of splits, $s$, affect both the bound and computation. The runtime follows, as expected, an $s^2$ time complexity and the bound does tighten with increasing $s$. Note, the `enhancement' approach of rerunning the algorithm with more slices on the most sensitive dimensions provides a more efficient way of using compute. We ran the same analysis on the credit dataset achieving 80\% accuracy (chance 50\%). We found that we started needing two inputs to be perturbed when we reached six slices, but were not close to reaching a bound of three (even with 100 slices). See supplementary for details of these results.
\subsection{Effect of using a sparse approximation}
We also investigated the effect of the use of a sparse approximation on the results. Using the $8 \times 8$ MNIST data (3 vs 5, 1000 training points, lengthscale 1, kernel variance of 1). Figure \ref{sparseandsliceeffect} illustrates the effect of increasing the number of inducing inputs. First, the accuracy and computation time increase, as one would expect. The algorithm's upper bound on the change in the posterior increases (weakens) with more inducing points. Increasing the number of EQs leads to a looser bound directly, but also means they become closer together, leading to steeper gradients in the posterior. With just two inducing points, at least three inputs need to be altered to cause a confident misclassification (but the classifier has only a 78\% accuracy). With five inducing points the accuracy is over 90\% but now the algorithm only guarantees two inputs need to be modified. In summary, a sparse approximation is a useful way to improve the algorithm's guarantee on the robustness. We investigated the same question with the credit dataset. With more than four inducing points the classifier was bound such that only one input needed to change. With two inducing inputs it is close to needing three inputs to change. See supplementary for details of these results.
\subsection{Bounds provided by Empirical Attack}
The AB algorithm provides a lower bound on the number of dimensions that need perturbing to cause a confident misclassification. While an empirical attack result provides an upper bound. 
\citet{papernot2016limitations} find the gradient of the prediction with respect to each input, and alter the inputs with the largest gradients first.
For GPC however, this often placed the example away from the training data, causing the posterior to return to the prior mean making us unable to reach the 95\% threshold using the JSM algorithm.
The original paper targeted a neural network with sigmoid activation functions. Both the linear and saturation aspects mean that it is likely that assigning the extreme value to these pixels will successfully generate the adversarial example required, while the GP has a less monotonic relationship with its inputs.
%
\begin{figure}
  \centering
  \includegraphics[width=0.47\textwidth]{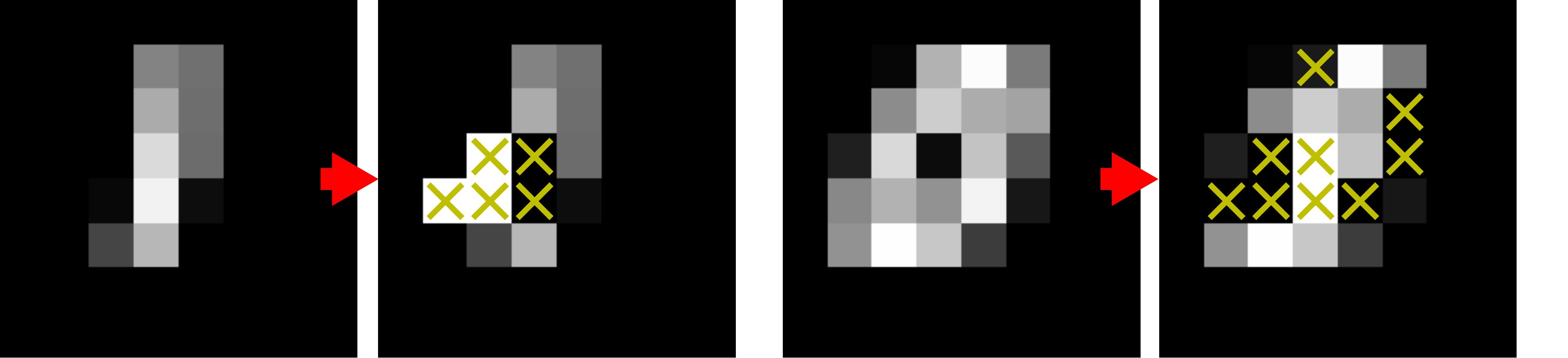}
\caption{Demonstration of perturbation necessary from a confidently classified one, left; or zero, right; to become a confidently classified zero or one respectively. Crosses, pixels that were modified. Initial images are from test set.}
  \label{singleexamples}
\end{figure}
%
Instead, we still choose the input with the largest absolute gradient but then performed a search along this input's axis. Figure \ref{singleexamples} gives two examples of such a perturbation. 
We applied this algorithm to all 470 confidently classified test points. We found examples in which just five pixels needed changing, providing an upper bound. The AB algorithm was applied to this model (lengthscale = 4.0, kernel variance = 202, both optimised using maximum likelihood). This found a lower bound of 3 pixels needing to be modified. Thus we know the true bound is between 3 and 5 (inclusive).

We briefly experimented with the algorithm described in \citet{carlini2017towards} for the $L_0$ attack, but found this produced looser bounds than the above approach. That algorithm chooses pixels to remove from the set by using the gradient of the GP latent function at the perturbed point, but it could be that this isn't a good method for GPC, especially at a target threshold far from the 50\% decision boundary.

\section{Discussion}

We have developed a method that provides guaranteed limits on the ability of an attacker to cause a confident classification to become a confident misclassification. 
This improves on previous efforts to evaluate the potential for adversarial attacks, as this method holds for the whole domain, and not just a localised volume around each training point.

Several interesting features are apparent. For most of the datasets, more regularisation appears to lead to more robust classifiers (i.e. more pixels need to change to cause a confident change in classification). This is largely expected but probably depends on the type of regularisation used \cite{grosse2018killing,demontis2017yes}.  There is an exception to this rule in the banknote dataset, where we see very short lengthscales also lead to robustness. 
With long-lengthscales it is, in effect, behaving much like LR, with almost-linear responses. At very short lengthscales it may or may-not be robust, depending on where the training data lies. This reflects the findings of \citet{grosse2018killing} who show empirically that both long and short lengthscales for a GP classifier can provide robustness against some attacks.

The algorithm scales to 100s of input dimensions (a typical limit of a GPC itself). We considered examples of datasets that require protection from adversarial attack (credit-worthiness, spam-classification and banknote-forgery). The use of a sparse approximation was found to improve the classifier's robustness bound, and allows larger training sets to be used.

For future work, extending this to a deep framework such as a deep GP \cite{damianou2013deep} may be surprisingly straightforward. The current AB algorithm allows one to bound the number of variables, for a given layer, required to cause a specific change in the layer's output. With a slight modification one could consider the combinations of all paths in which each dimension has a limited perturbation. Although potentially computationally intractable for high dimensionality, typical deep GPs have few dimensions beyond the input layer \cite{damianou2013deep}.

We decided against using the uncertainty estimates that the GP classifier provides. It has been found \cite{grosse2018limitations} that this uncertainty can assist in the detection of off-the-shelf adversarial attacks. However, the latent mean is usually near zero in uncertain locations anyway (for reasonable lengthscales and the EQ kernel) so this would be of limited benefit.

Future computational improvements include the use of dynamic programming when summing path segments and iterative refinement of the grid when finding the bound on the mixture of EQs. Other improvements to the actual value of the bound could be achieved. For example in how the pairs of positive and negative EQs are selected when cancelling out the negative terms (currently the nearest positive EQ is greedily chosen). Other kernels could be used by adjusting the calculation of the bounded mixture of kernels. 

We believe that it is vital we can construct classifiers, confident that imperceptibly small changes will not cause large changes in classification. The presence of such vulnerabilities in classifiers used in safety critical applications is of concern. Our method is a first step towards alleviating this. In particular is the danger of `blind spot attacks' as described in \citet{zhang2019limitations}. These are overlooked by the common certification methods that just work around the epsilon ball.

How one uses the bound to construct such classifiers is a more open question. We imagine that the designer or user of the classifier could run our algorithm on their classifier and training data and select model parameters to trade off the accuracy against this bound.

\subsubsection{Conclusion}

In summary, we have devised a framework for bounding the scale of an $L_0$ adversarial attack leading to a confident misclassification, developed a method for applying the bound to GP classification and through a series of experiments tested this bound against similarly bounded LR and investigated the effect of classifier configuration parameters. We found our GP classifier could be both accurate and provably robust. This is the first paper to successfully find such a bound. This method provides a foundation for future research - in particular for expanding to larger datasets and more complex classifiers.

\subsection{Acknowledgments}
Funded by the EPSRC (EP/N014162/1) and the BMBF (FKZ: 16KIS0753).

\clearpage
\bibliographystyle{plainnat}
\bibliography{refs}

\begin{thebibliography}{28}
\providecommand{\natexlab}[1]{#1}
\providecommand{\url}[1]{\texttt{#1}}
\expandafter\ifx\csname urlstyle\endcsname\relax
  \providecommand{\doi}[1]{doi: #1}\else
  \providecommand{\doi}{doi: \begingroup \urlstyle{rm}\Url}\fi

\bibitem[Biggio et~al.(2013)Biggio, Corona, Maiorca, Nelson, Srndic, Laskov,
  Giacinto, and Roli]{BiggioCMNSLGR13}
Battista Biggio, Igino Corona, Davide Maiorca, Blaine Nelson, Nedim Srndic,
  Pavel Laskov, Giorgio Giacinto, and Fabio Roli.
\newblock Evasion attacks against machine learning at test time.
\newblock In \emph{Machine Learning and Knowledge Discovery in Databases -
  European Conference, {ECML} {PKDD} 2013, Prague, Czech Republic, September
  23-27, 2013, Proceedings, Part {III}}, pages 387--402, 2013.

\bibitem[Cardelli et~al.(2019)Cardelli, Kwiatkowska, Laurenti, and
  Patane]{cardelli2019robustness}
Luca Cardelli, Marta Kwiatkowska, Luca Laurenti, and Andrea Patane.
\newblock Robustness guarantees for {B}ayesian inference with {G}aussian
  processes.
\newblock In \emph{Proceedings of the AAAI Conference on Artificial
  Intelligence}, volume~33, pages 7759--7768, 2019.

\bibitem[Carlini and Wagner(2017{\natexlab{a}})]{carlini2017adversarial}
Nicholas Carlini and David Wagner.
\newblock Adversarial examples are not easily detected: Bypassing ten detection
  methods.
\newblock In \emph{Proceedings of the 10th ACM Workshop on Artificial
  Intelligence and Security}, pages 3--14. ACM, 2017{\natexlab{a}}.

\bibitem[Carlini and Wagner(2017{\natexlab{b}})]{carlini2017towards}
Nicholas Carlini and David Wagner.
\newblock Towards evaluating the robustness of neural networks.
\newblock In \emph{2017 IEEE Symposium on Security and Privacy (SP)}, pages
  39--57. IEEE, 2017{\natexlab{b}}.

\bibitem[Carlini et~al.(2017)Carlini, Katz, Barrett, and
  Dill]{carlini2017provably}
Nicholas Carlini, Guy Katz, Clark Barrett, and David~L Dill.
\newblock Provably minimally-distorted adversarial examples.
\newblock \emph{arXiv preprint arXiv:1709.10207}, 2017.

\bibitem[Carreira-Perpinan(2000)]{carreira2000mode}
Miguel~A. Carreira-Perpinan.
\newblock Mode-finding for mixtures of {G}aussian distributions.
\newblock \emph{IEEE Transactions on Pattern Analysis and Machine
  Intelligence}, 22\penalty0 (11):\penalty0 1318--1323, 2000.

\bibitem[Cranor and LaMacchia(1998)]{cranor1998spam}
Lorrie~Faith Cranor and Brian~A LaMacchia.
\newblock Spam!
\newblock \emph{Communications of the ACM}, 41\penalty0 (8):\penalty0 74--83,
  1998.

\bibitem[Damianou and Lawrence(2013)]{damianou2013deep}
Andreas Damianou and Neil Lawrence.
\newblock Deep {G}aussian processes.
\newblock In \emph{Artificial Intelligence and Statistics}, pages 207--215,
  2013.

\bibitem[Demontis et~al.(2017)Demontis, Melis, Biggio, Maiorca, Arp, Rieck,
  Corona, Giacinto, and Roli]{demontis2017yes}
Ambra Demontis, Marco Melis, Battista Biggio, Davide Maiorca, Daniel Arp,
  Konrad Rieck, Igino Corona, Giorgio Giacinto, and Fabio Roli.
\newblock Yes, machine learning can be more secure! {A} case study on {A}ndroid
  malware detection.
\newblock \emph{IEEE Transactions on Dependable and Secure Computing}, 2017.

\bibitem[Dua and Graff(2017)]{asuncion2007uci}
Dheeru Dua and Casey Graff.
\newblock {UCI} machine learning repository, 2017.
\newblock URL \url{http://archive.ics.uci.edu/ml}.

\bibitem[Grosse et~al.(2018{\natexlab{a}})Grosse, Pfaff, Smith, and
  Backes]{grosse2018limitations}
Kathrin Grosse, David Pfaff, Michael~T Smith, and Michael Backes.
\newblock The limitations of model uncertainty in adversarial settings.
\newblock \emph{arXiv preprint arXiv:1812.02606}, 2018{\natexlab{a}}.

\bibitem[Grosse et~al.(2018{\natexlab{b}})Grosse, Smith, and
  Backes]{grosse2018killing}
Kathrin Grosse, Michael~T Smith, and Michael Backes.
\newblock Killing four birds with one {G}aussian process: Analyzing test-time
  attack vectors on classification.
\newblock \emph{arXiv preprint arXiv:1806.02032}, 2018{\natexlab{b}}.

\bibitem[Hein and Andriushchenko(2017)]{hein2017formal}
Matthias Hein and Maksym Andriushchenko.
\newblock Formal guarantees on the robustness of a classifier against
  adversarial manipulation.
\newblock In \emph{Advances in Neural Information Processing Systems}, pages
  2266--2276, 2017.

\bibitem[Huang et~al.(2017)Huang, Kwiatkowska, Wang, and Wu]{huang2017safety}
Xiaowei Huang, Marta Kwiatkowska, Sen Wang, and Min Wu.
\newblock Safety verification of deep neural networks.
\newblock In \emph{International Conference on Computer Aided Verification},
  pages 3--29. Springer, 2017.

\bibitem[Krizhevsky et~al.(2012)Krizhevsky, Sutskever, and
  Hinton]{NIPS20124824}
Alex Krizhevsky, Ilya Sutskever, and Geoffrey~E Hinton.
\newblock Imagenet classification with deep convolutional neural networks.
\newblock In F.~Pereira, C.~J.~C. Burges, L.~Bottou, and K.~Q. Weinberger,
  editors, \emph{Advances in Neural Information Processing Systems 25}, pages
  1097--1105. Curran Associates, Inc., 2012.

\bibitem[Madry et~al.(2018)Madry, Makelov, Schmidt, Tsipras, and
  Vladu]{madry2018towards}
Aleksander Madry, Aleksandar Makelov, Ludwig Schmidt, Dimitris Tsipras, and
  Adrian Vladu.
\newblock Towards deep learning models resistant to adversarial attacks.
\newblock In \emph{6th International Conference on Learning Representations,
  {ICLR} 2018, Vancouver, BC, Canada, April 30 - May 3, 2018, Conference Track
  Proceedings}, 2018.

\bibitem[Papernot et~al.(2016{\natexlab{a}})Papernot, McDaniel, Jha,
  Fredrikson, Celik, and Swami]{papernot2016limitations}
Nicolas Papernot, Patrick McDaniel, Somesh Jha, Matt Fredrikson, Z~Berkay
  Celik, and Ananthram Swami.
\newblock The limitations of deep learning in adversarial settings.
\newblock In \emph{Security and Privacy (EuroS\&P), 2016 IEEE European
  Symposium on}, pages 372--387. IEEE, 2016{\natexlab{a}}.

\bibitem[Papernot et~al.(2016{\natexlab{b}})Papernot, McDaniel, Wu, Jha, and
  Swami]{papernot2016distillation}
Nicolas Papernot, Patrick McDaniel, Xi~Wu, Somesh Jha, and Ananthram Swami.
\newblock Distillation as a defense to adversarial perturbations against deep
  neural networks.
\newblock In \emph{2016 IEEE Symposium on Security and Privacy (SP)}, pages
  582--597. IEEE, 2016{\natexlab{b}}.

\bibitem[Peck et~al.(2017)Peck, Roels, Goossens, and Saeys]{peck2017lower}
Jonathan Peck, Joris Roels, Bart Goossens, and Yvan Saeys.
\newblock Lower bounds on the robustness to adversarial perturbations.
\newblock In \emph{Advances in Neural Information Processing Systems}, pages
  804--813, 2017.

\bibitem[Pulkkinen et~al.(2013)Pulkkinen, M{\"a}kel{\"a}, and
  Karmitsa]{pulkkinen2013continuation}
Seppo Pulkkinen, Marko~Mikael M{\"a}kel{\"a}, and Napsu Karmitsa.
\newblock A continuation approach to mode-finding of multivariate {G}aussian
  mixtures and kernel density estimates.
\newblock \emph{Journal of Global Optimization}, 56\penalty0 (2):\penalty0
  459--487, 2013.

\bibitem[Ross and Doshi-Velez(2018)]{ross2018improving}
Andrew~Slavin Ross and Finale Doshi-Velez.
\newblock Improving the adversarial robustness and interpretability of deep
  neural networks by regularizing their input gradients.
\newblock In \emph{Thirty-Second AAAI Conference on Artificial Intelligence},
  2018.

\bibitem[Sitawarin et~al.(2018)Sitawarin, Bhagoji, Mosenia, Mittal, and
  Chiang]{sitawarin2018rogue}
Chawin Sitawarin, Arjun~Nitin Bhagoji, Arsalan Mosenia, Prateek Mittal, and
  Mung Chiang.
\newblock Rogue signs: Deceiving traffic sign recognition with malicious ads
  and logos.
\newblock \emph{arXiv preprint arXiv:1801.02780}, 2018.

\bibitem[Snelson and Ghahramani(2006)]{snelson2006sparse}
Edward Snelson and Zoubin Ghahramani.
\newblock Sparse {G}aussian processes using pseudo-inputs.
\newblock In \emph{Advances in neural information processing systems}, pages
  1257--1264, 2006.

\bibitem[Su et~al.(2019)Su, Vargas, and Sakurai]{su2019one}
Jiawei Su, Danilo~Vasconcellos Vargas, and Kouichi Sakurai.
\newblock One pixel attack for fooling deep neural networks.
\newblock \emph{IEEE Transactions on Evolutionary Computation}, 2019.

\bibitem[Szegedy et~al.(2014)Szegedy, Zaremba, Sutskever, Bruna, Erhan,
  Goodfellow, and Fergus]{szegedy2013intriguing}
Christian Szegedy, Wojciech Zaremba, Ilya Sutskever, Joan Bruna, Dumitru Erhan,
  Ian Goodfellow, and Rob Fergus.
\newblock Intriguing properties of neural networks.
\newblock \emph{ICLR}, 2014.

\bibitem[Williams and Rasmussen(2006)]{williams2006gaussian}
Christopher~KI Williams and Carl~Edward Rasmussen.
\newblock Gaussian processes for machine learning.
\newblock \emph{MIT Press}, 2006.

\bibitem[Wong and Kolter(2018)]{wong2018provable}
Eric Wong and Zico Kolter.
\newblock Provable defenses against adversarial examples via the convex outer
  adversarial polytope.
\newblock In \emph{International Conference on Machine Learning}, pages
  5283--5292, 2018.

\bibitem[Zhang et~al.(2019)Zhang, Chen, Song, Boning, Dhillon, and
  Hsieh]{zhang2019limitations}
Huan Zhang, Hongge Chen, Zhao Song, Duane~S. Boning, Inderjit~S. Dhillon, and
  Cho{-}Jui Hsieh.
\newblock The limitations of adversarial training and the blind-spot attack.
\newblock In \emph{7th International Conference on Learning Representations,
  {ICLR} 2019, New Orleans, LA, USA, May 6-9, 2019}, 2019.

\end{thebibliography}

\clearpage
\section*{Supplemental Material}

\subsection*{Bounding the sum of a mixture of exponentiated quadratics (EQs)}
\label{mixbound}

We need to be able to bound the sum of a mixture of exponentiated quadratics (EQs). Research exists looking at heuristics and methods for approximating this sum  \cite{carreira2000mode,pulkkinen2013continuation} but we are interested in finding a strict bound on the peak. Specific to our problem, each EQ has the same, isotropic, covariance, but the contribution weights for the EQs can be negative. We first consider the case of the low-($d$)-dimensional input with positive only weights which we can exhaustively search with an evenly $s$-spaced grid and will later provide methods for handling higher dimensions and negative weights. We can evaluate the mixture at each grid point but the actual peak is almost certain to lie between grid points. The furthest distance from a grid point is $\frac{1}{2} s \sqrt{d}$. The worst case is that this consists of a Gaussian centred at that furthest point. Thus we assume this worst case contingency and assume that the actual peak is equal to the maximum grid value divided by $\exp\left[-\frac{(\frac{1}{2}s\sqrt{d})^2}{2 l^2} \right]$, the EQ kernel function with lengthscale $l$.
\begin{figure}[h!]
  \begin{center}
  \includegraphics[width=0.45\textwidth,trim=40 50 10 60,clip]{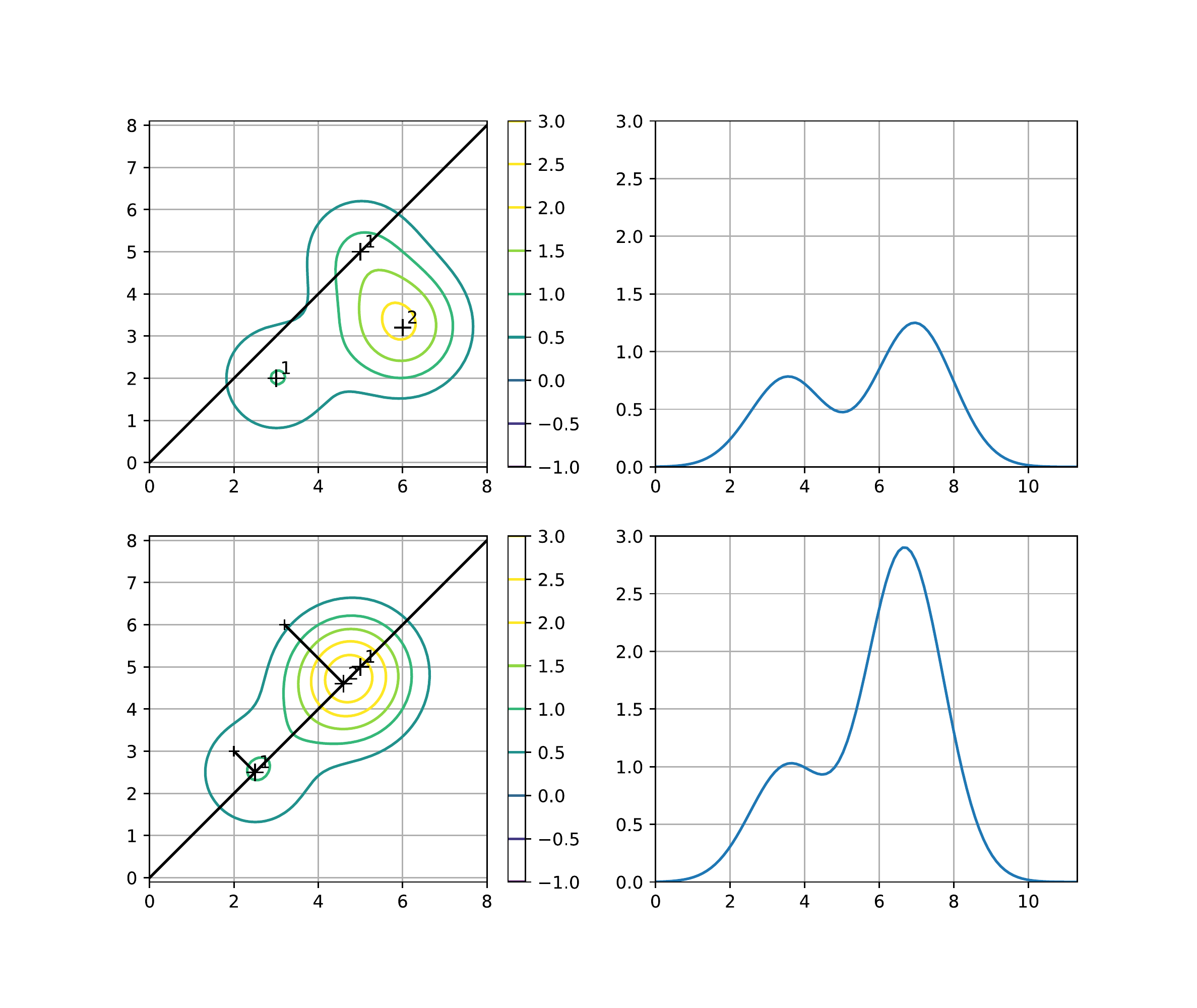}
  \end{center}
\caption{Upper left, contour plot of the sum of three, two-dimensional EQs. Upper right, the values of the sum along the diagonal, one-dimensional line. Lower left, the effect on the contour plot if the three points are placed on the diagonal line to illustrate the effect of a principle component approximation. Lower right, the sum along the diagonal line. Note that this new function is never less than the function for when the EQ points are spaced in the full two-dimensional domain. \label{lowrank}}
\end{figure}
We now consider the high dimensional situation (again with just positive weights). We simply perform PCA on the locations of the EQs and then apply the simple grid algorithm to the low-dimensional manifold. One can see that any bound on the sum of inputs for the lower-dimensional manifold will hold for the full domain (assuming positive weights, see figure \ref{lowrank}). To prove this, first consider how the distance between two points will differ in the $k$-(low)-dimensional manifold vs in the ($d$-dimensional) full-domain. We can factorise the PCA transformation matrix, $\bm{W}$, into a rotation, $\bm{R}$, and a simple dimension removal matrix, $\bm{B}$. Without loss of generality we can chose the rotation to ensure that it is the first $1..k$ dimensions which are preserved (i.e. so $\bm{B}$ is a $d \times k$ rectangular matrix, consisting of the $k \times k$ identity matrix in the top sub matrix and zeros in the bottom sub matrix). The distance between any pair of points is invariant to the rotation. We take a pair of points, $\bm{x}_1 \bm{R}$ and $\bm{x}_2 \bm{R}$, from after the rotation. The distance between them in the full domain is $||\bm{x}_1 \bm{R}-\bm{x}_2 \bm{R}||_2 = \sqrt{\sum_{i=1}^d[(\bm{x}_1-\bm{x}_2) \bm{R}]_{(i)}^2}$ and in the low-dimensional hyperplane $||(\bm{x}_1-\bm{x}_2) \bm{R} \bm{B}||_2 = \sqrt{\sum_{i=1}^k[(\bm{x}_1-\bm{x}_2) \bm{R}]_{(i)}^2}$. Where the $\bm{B}$ is not included as it is represented by only summing the $k$ included dimensions. We note that the terms the sums in the two equations are all non-negative (as they consist of the square of real numbers) and so the sum over the $k$ dimensions will necessarily not be less than the sum over the $d \geq k$ dimensions. We also note that the EQ decreases monotonically with increasing distance. For any location $\bm{x}_*$ the distances to any training point $\bm{x}_i$ will not increase when transformed to the $k$-dimensional manifold, and thus the associated EQ contribution to the weighted sum at $\bm{x}_*$ will not be reduced. As this is true for all test points and over all training points, we can see that the maximum of the weighted sum of EQs can not get smaller. Thus if we find an upper bound on the sum of the $k$-dimensional domain this will also hold for the full domain. There are no guarantees that the bound will be a close one, although, if the $k$ principle components used capture sufficient variance in the data the residuals will not contribute as much to the distance between points.

We finally need to consider the negatively weighted EQ bases. One could set these to zero, and accept a looser bound. However, we found for this application there was a more efficient way to treat them.

Consider a one dimensional function $f(x)$ made by summing a positively weighted and a negatively weighted EQ (of equal lengthscale), placed at the origin and at 1, respectively. We assume the function has positive values and a maximum $y_0$, at $x_0$. We propose that one can replace this summed pair of EQs with a function $f'(x)$ consisting of a single positive EQ, located at $x_0$ and with weight $y_0$. The new function $f'$ upper bounds the original function, $f$. We can extend this proof to higher dimensions by considering this one dimension as lying on the line between two EQ centres in a high dimensional domain. Every parallel line in the domain has the same functions (but scaled) $w f(x)$ and $w f'(x)$. As the two functions are scaled equally the same bounds will apply.

\subsection*{Method for combining the sum of positive and negative EQs}
\begin{figure}
  \begin{center}
  \includegraphics[width=0.45\textwidth]{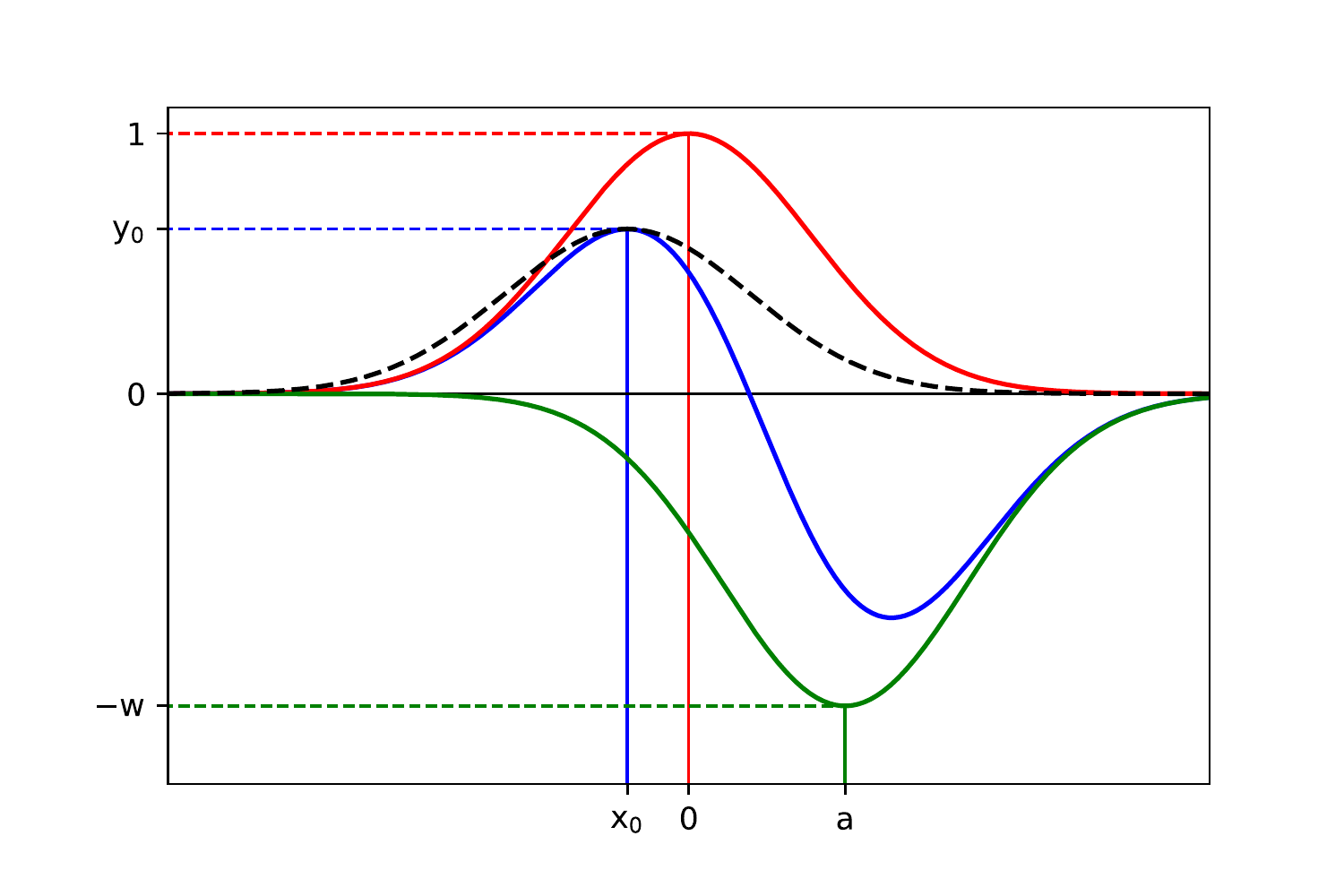}
  \end{center}
\caption{We believe that a single EQ placed and scaled to the maximum of the sum of positive and negative EQs bounds the sum.}
\end{figure}

To tighten the bound we wish to incorporate the negative-weighted EQs, not just discard them. To this end we wish to show that a function $f(x)$ equal to the sum of two EQs (exponentiated quadratics) is upper bounded by a single EQ, placed at the peak of $f(x)$, i.e;
\begin{equation}
\label{sumeq}
f(x) = e^{-x^2} - w e^{-(x-a)^2} \leq y_0 e^{-(x-x_0)^2}
\end{equation}
Where $y_0$ and $x_0$ refer to the value and location of the maximum of $f(x)$. $a$ specifies the location of the negative EQ (and $a$ is defined to be non-negative for this proof). $w$ specifies the scale of the negative term. We assume for this proof that $a$ and $w$ are chosen such that $y_0>0$. We proceed as follows. We first note that the gradient of $f(x)$ should be zero (wrt $x$) at the maximum (where $x=x_0$),

\begin{equation}
\frac{df}{dx} = -2 x e^{-x^2} + 2(x-a)w e^{-(x-a)^2} = 0.
\end{equation}
So,
\begin{equation}
2 x_0 e^{-x_0^2} = 2(x_0-a)w e^{-(x_0-a)^2}.
\end{equation}
This gives us an expression for $w$,
\begin{equation}
w = \frac{x_0}{x_0-a} e^{a^2-2 x_0 a}.
\end{equation}
We note that $y_0$ is simply equation \ref{sumeq} evaluated at $x_0$, so $y_0 = e^{-x_0^2} - w e^{-(x_0-a)^2}$.

Substituting in our expressions for $w$ and $y_0$ into inequality \ref{sumeq} we have,
\begin{equation}
\label{sumeqfull}
e^{-x^2} - \frac{x_0}{x_0-a} e^{a^2-2 x_0 a} e^{-(x-a)^2} \leq \left[ e^{-x_0^2} - \frac{x_0}{x_0-a} e^{a^2-2 x_0 a} e^{-(x_0-a)^2}\right] e^{-(x-x_0)^2}
\end{equation}
and multiplying out the bracket,
\begin{equation}
e^{-x^2} - \frac{x_0}{x_0-a} e^{a^2-2 x_0 a - x^2 +2ax - a^2} \leq e^{-x_0^2-x^2+2 x x_0 -x_0^2} - \frac{x_0}{x_0-a} e^{a^2-2 x_0 a -x_0^2 +2x_0 a -a^2-x^2+2 x x_0 -x_0^2}.
\end{equation}

Dividing both sides by $\frac{x_0}{x_0-a} e^{-x^2}$ (this is always positive), and cancelling some exponential terms, we are left with;

\begin{equation}
\frac{x_0-a}{x_0} - e^{-2 x_0 a +2ax} \leq \left(\frac{x_0-a}{x_0}-1\right) e^{-2x_0^2+2 x x_0}
\end{equation}
Which finally results in,

\begin{equation}
1 - \frac{a}{x_0} - e^{2 x a-2 x_0 a} \leq -\frac{a}{x_0} e^{2 x x_0-2x_0^2}.
\end{equation}
We define a function $g(x)$ which is simply the result of subtracting the right hand side of the inequality from the left, 
\begin{equation}
\label{g}
g(x) = 1 - \frac{a}{x_0} - e^{2 x a-2 x_0 a} + \frac{a}{x_0} e^{2 x x_0-2x_0^2} \leq 0.
\end{equation}
We wish to show that $g(x)$ is never greater than zero. In the steps that follow, we shall show that it has only two turning points at $x_0$ and $a$; we will show that $g(x)$ only has one (finite) point where it equals zero; we will show that this location is at $x_0$; we show that this turning point at $x_0$ is a maximum; finally we confirm that at this maximum $g(x)$ is non-positive; and thus as it is the only location where $g(x)=0$, the function must be negative everywhere else, and so $g(x)\leq 0$.

First we note that $g(x)$'s derivative wrt $x$ has only two zero-crossing points (for finite $x$). The gradient can be written as
$\frac{dg}{dx} = -2 a e^{-2 a x_0} e^{2 x a} + 2 a e^{-2 x_0^2} e^{2 x x_0}$. This expression equals zero (for non-zero $a$) in two cases. One at $x=a$ and one where $x=x_0$. The function $g(x)$ only equals zero for one (finite) value of $x$. Setting the expression for $g(x)=0$ in equation \ref{g}, then rearranging and solving for $x$ gives us $x=x_0$. Thus the turning point at $x=x_0$ is the only (finite) location that $g(x)=0$ too. We can disregard the other turning point. We now simply must show that $g(x)$ is at a maximum at $x=x_0$ (and thus $g(x) \leq g(x_0)$ for all $x$). We do this by differentiating again, $\frac{d^2g}{dx^2} = -4 a^2 e^{-2 a x_0} e^{2 x a} + 4 a x_0 e^{-2 x_0^2} e^{2 x x_0}$. Setting $x=x_0$, $\frac{d^2g}{dx^2} = 4 a (x_0 - a)$. This expression is never positive (as $x_0$ is never positive and $a$ is non-negative). So this is a maximum location. We note again that using equation \ref{g} we find that $g(x_0)=0$. As this is a maximum and the only location where $g(x)=0$ we can state that $g(x) \leq 0$ for all finite values of $x$. Thus our original inequality \ref{sumeq} holds too, i.e. an EQ function of scale $y_0$ at location $x_0$ is never less than $f(x)$.
\begin{table*}
\begin{center}
\begin{tabular}{r p{2.6cm} r r r r l}
Slices & Number of inputs to change & Elapse time (s) & \multicolumn{4}{r}{Cumulative sum of top four dimensions} \\
\hline
1 & 1 & 49 & 2.48 & 4.96 & 7.32 & 9.61\\
2 & 1 & 46 & 2.46 & 4.88 & 7.22 & 9.45\\
3 & 1 & 46 & 2.34 & 4.48 & 6.57 & 8.62\\
4 & 1 & 44 & 2.23 & 4.32 & 6.23 & 8.08\\
6 & 2 & 53 & 2.08 & 4.09 & 5.90 & 7.69\\
8 & 2 & 53 & 2.00 & 3.98 & 5.73 & 7.45\\
10 & 2 & 53 & 1.97 & 3.92 & 5.63 & 7.33\\
12 & 2 & 50 & 1.95 & 3.87 & 5.57 & 7.25\\
15 & 2 & 53 & 1.94 & 3.83 & 5.50 & 7.17\\
20 & 2 & 58 & 1.92 & 3.78 & 5.44 & 7.09\\
30 & 2 & 73 & 1.90 & 3.74 & 5.37 & 7.01\\
50 & 2 & 126 & 1.89 & 3.71 & 5.33 & 6.94\\
70 & 2 & 190 & 1.88 & 3.69 & 5.30 & 6.91\\
100 & 2 & 335 & 1.88 & 3.68 & 5.29 & 6.89\\
\hline
\end{tabular}
\end{center}
\caption{Credit dataset: Effect of increasing the number of slices the domain is divided on the lower bound on the number of inputs required to cause a confident misclassification and on computation time. The cumulative effect of the first, two, three and fourth most significant input dimensions are also listed. The distance between the 5th and 95th percentile training points in the posterior was 2.132.}
\label{effectofsplitscredit}
\end{table*}
\subsection*{Additional Adversarial Examples}
\begin{figure}
  \begin{center}
  \includegraphics[width=0.15\textwidth,trim=10 40 30 30,clip]{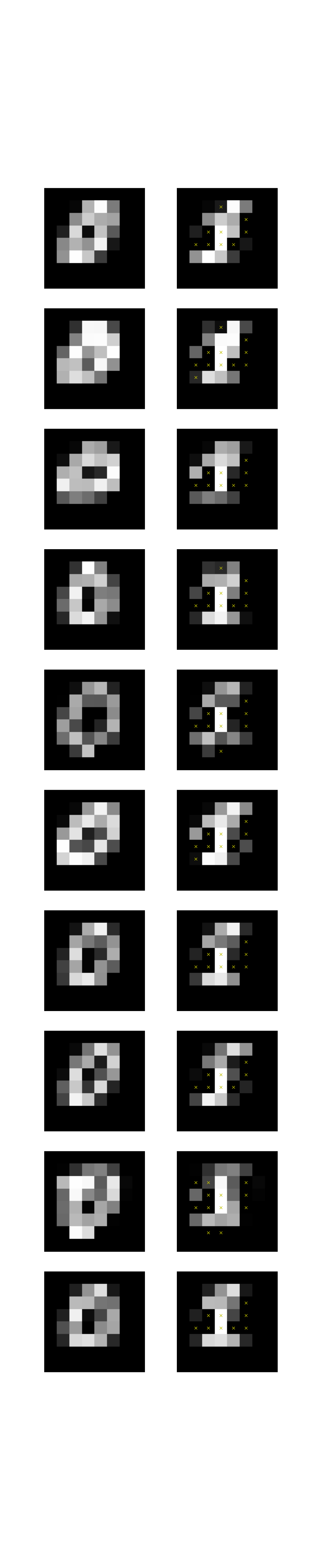}
  \includegraphics[width=0.15\textwidth,trim=10 40 30 30,clip]{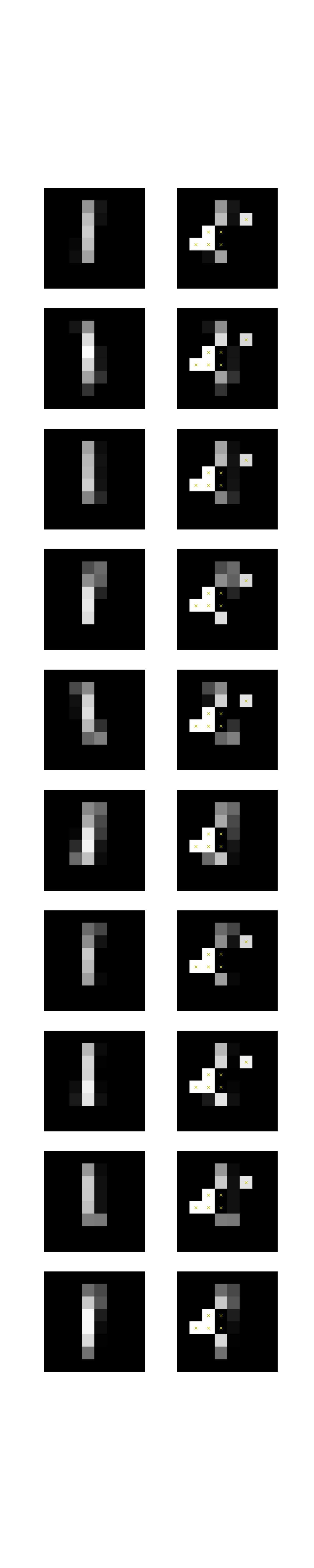}
  \end{center}
\caption{Demonstration of perturbation necessary from a confidently classified zero, left column pair; or one, right column pair; to become a confidently classified one or zero respectively. Crosses mark the pixels that were modified. Initial images are from test set. Greyscale from zero (black) to one (white).}
  \label{aeexamples}
\end{figure}

Figure \ref{aeexamples} demonstrates 10 adversarial examples (from each class).

\subsection*{$3 \times 3$ MNIST (low vs high)}
\begin{figure}
  \begin{center}
  \includegraphics[width=0.09\textwidth,trim=10 40 20 40,clip]{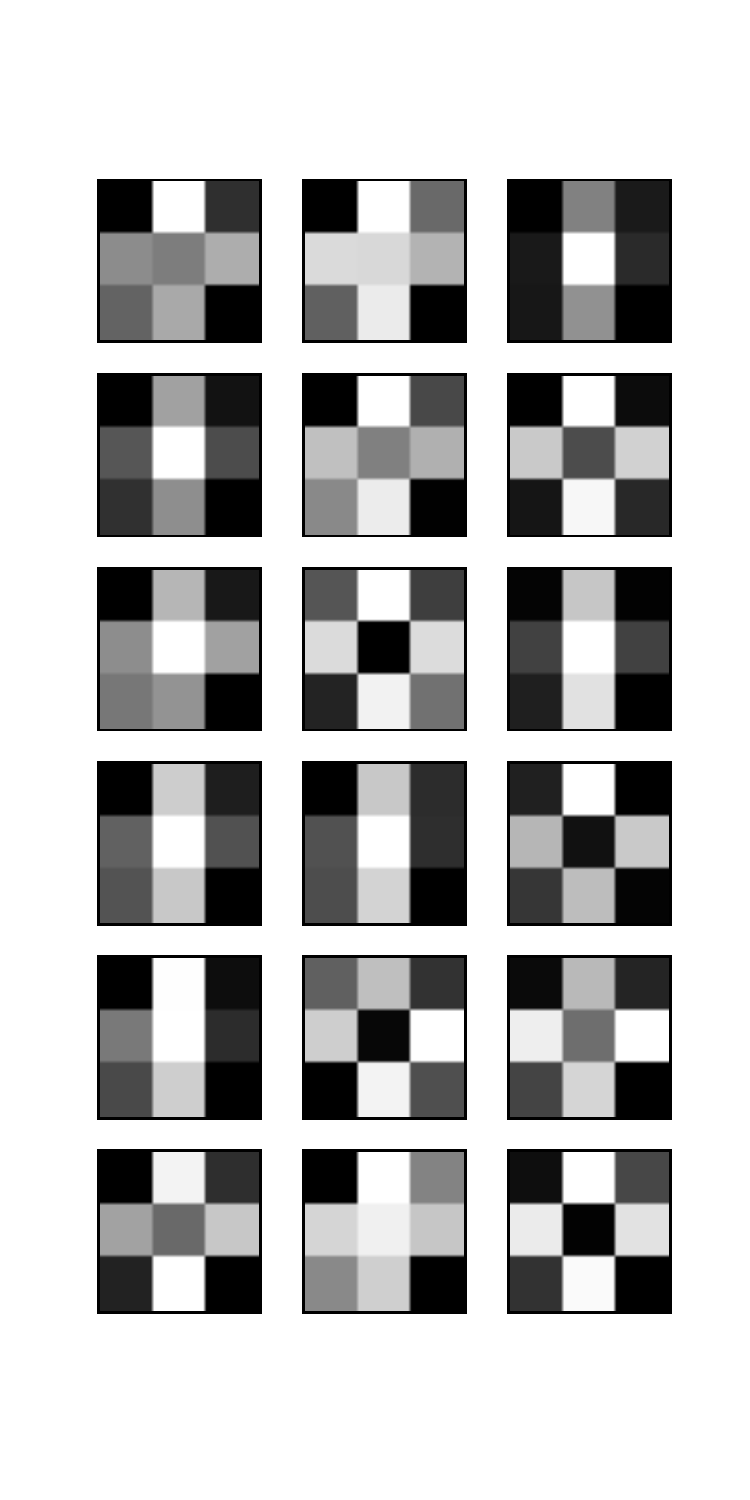}
  \includegraphics[width=0.09\textwidth,trim=10 40 20 40,clip]{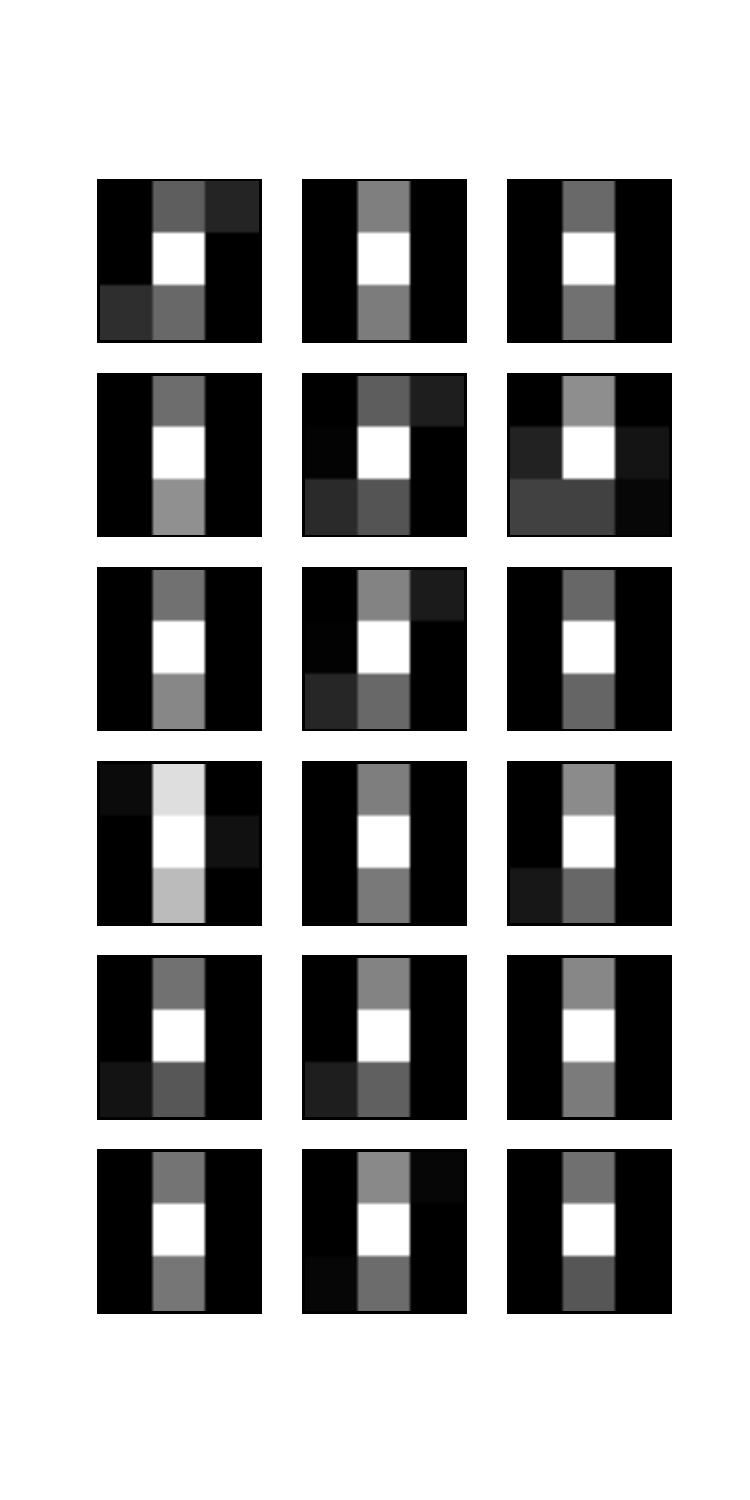}
  \includegraphics[width=0.09\textwidth,trim=10 40 20 40,clip]{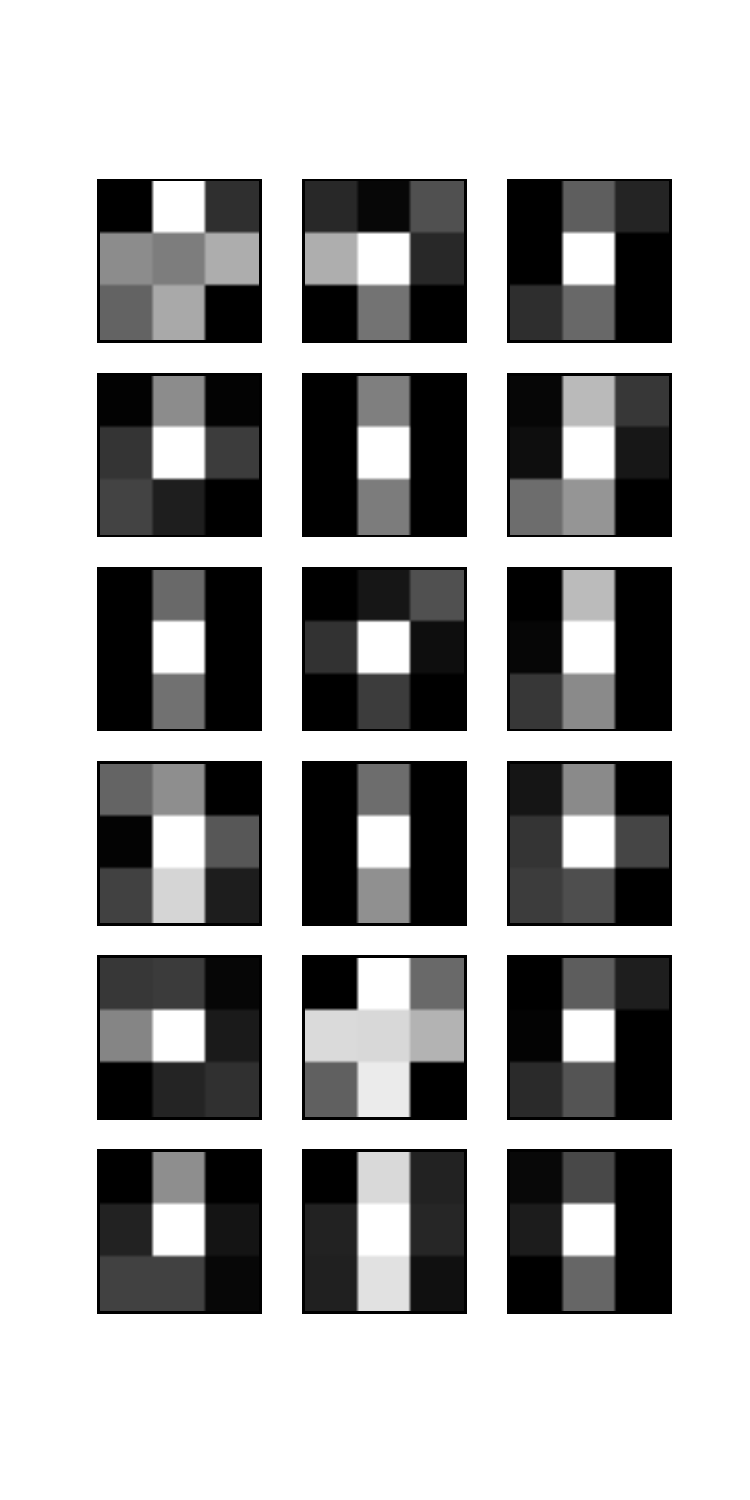}
  \includegraphics[width=0.09\textwidth,trim=10 40 20 40,clip]{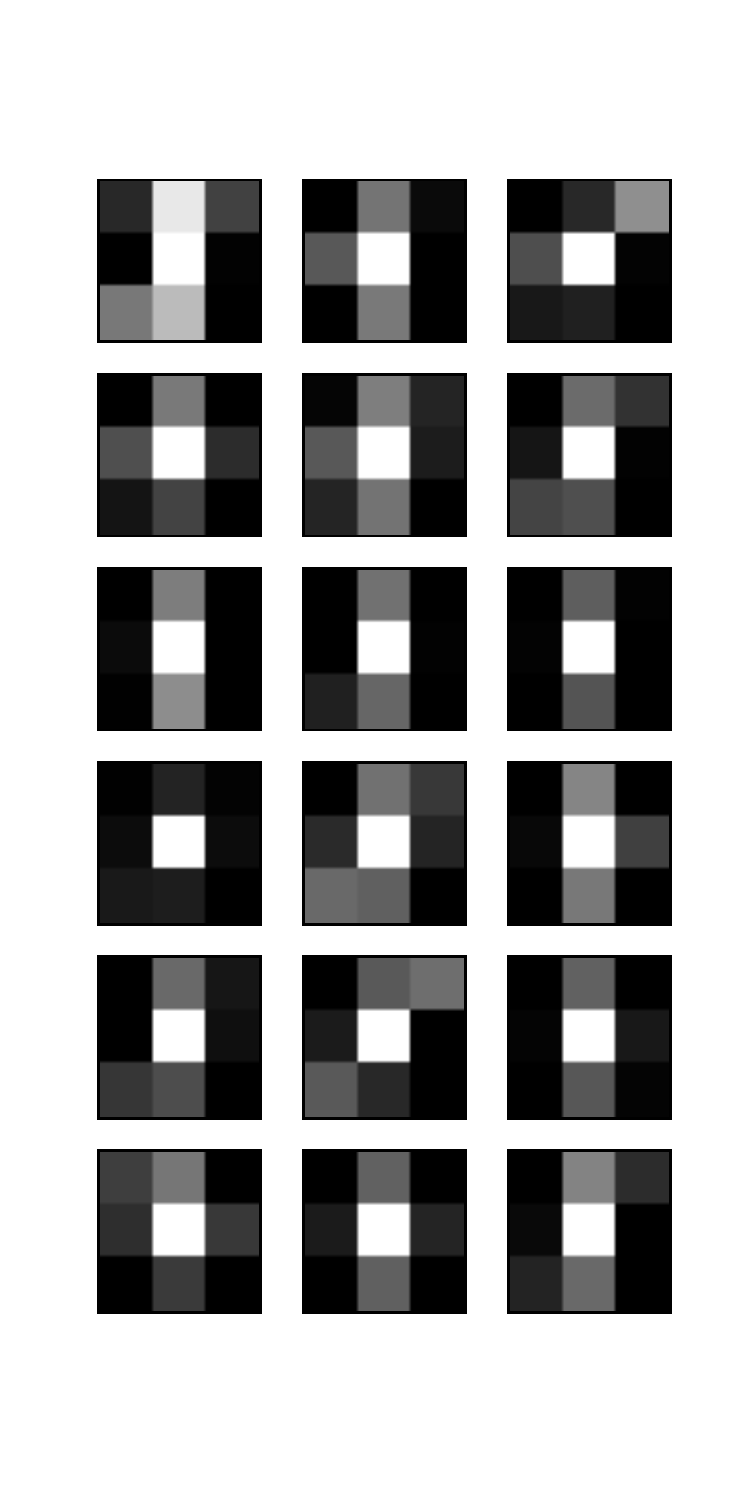}
  \end{center}
\caption{3x3 MNIST. Examples of the four class types used. From left-to-right; \{0\}, \{1\}, \{0..4\}, \{5..9\}.}
  \label{lowresmnistexamples}
\end{figure}
As a more concrete example we consider a $3 \times 3$ MNIST problem (Figure \ref{lowresmnistexamples}). 
Rather than just use a two digit classification problem which will probably be linearly separable, we classify digits as being from either the set of ${0,1,2,3,4}$ or ${5,6,7,8,9}$. A shorter lengthscale would typically be necessary as the decision boundary is unlikely to be a hyperplane (which a long lengthscale would approximate) but instead require a more complicated function. We normalised the data to lie inside the unit hypercube and used a lengthscale of one, kernel variance 0.12, $\sigma^2=0.01$ and we used 100 training points. The classifier achieved an accuracy of just 72\% (chance is 50\%), possibly due to the limited resolution. We split the domain into $2^9$ hypercubes to run the AB algorithm. We evaluated the latent function at all the training points and found it differed by 0.532 between the 5th and 95th percentiles.
We ran the AB algorithm and found an upper bound of 0.988 for one input change, far larger then than this threshold ($0.988>0.532$). One million randomly chosen one-input perturbations were tested. The most influential one only changed the latent function by 0.246, providing an empirical lower bound. There is a considerable gap between these two bounds (0.246 and 0.988), within which the distance between the two thresholds lies (0.532). We refine the search using the enhancement trick (using 24 steps in a higher resolution grid), which results in a bound of 0.509, just below the confident threshold (0.532). Hence to achieve a change from below the bottom 5th percentile to above the 95th, using this classifier will require at least \emph{two pixels} to be changed (one is provably not enough).
There are several parameters that need to be traded to achieve the tightest bound for a limited computational budget (e.g. the resolution and dimensionality of the PCA in the mixture-of-EQs-bound, the resolution of the enhancement, etc) which depend on the data and the lengthscale.
\begin{table*}
\begin{center}
\begin{tabular}{r r p{1.6cm} r p{2cm} r r r r}
Inducing Points & Accuracy & Number of inputs to change & Elapse time (s) & Threshold Distance & \multicolumn{4}{r}{Cumulative sum of top four dimensions} \\
\hline
2 & 0.79 & 2 & 208 & 2.28 & 1.62 & 2.86 & 4.08 & 5.05\\
4 & 0.80 & 2 & 335 & 2.13 & 1.88 & 3.68 & 5.29 & 6.89\\
6 & 0.82 & 1 & 477 & 1.94 & 2.64 & 5.06 & 7.15 & 8.99\\
8 & 0.82 & 1 & 497 & 1.44 & 23.22 & 42.95 & 62.53 & 79.32\\
10 & 0.82 & 1 & 717 & 1.33 & 70.77 & 135.76 & 182.26 & 221.20\\
\hline
\end{tabular}
\end{center}
\caption{Credit dataset: Effect of increasing the number of inducing points on accuracy, the lower bound on the number of inputs required to cause a confident misclassification and on computation time. The cumulative effect of the first, two, three and fourth most significant input dimensions are also listed. The threshold distance between the 5th and 95th percentile training points in the posterior is also recorded. Chance: 50\%.}
\label{effectofinducingscredit}
\end{table*}
If we briefly consider the case of simply classifying 0s and 1s, we find a higher number of pixel changes are required (according to the bound) to achieve a confident misclassification. Specifically, even using a (quicker) less detailed configuration we find that the two pixel bound is less than the confidence threshold, thus \emph{three} or more pixels need to change in order to move from a 5\% to a 95\% location. This increase in the bound is probably because the two classes are more clearly separated, making it hard to get between classes with a single pixel change. This is reflected in the higher accuracy (95\%) and is clear from considering examples of the four classes (figure \ref{lowresmnistexamples}).

\subsection*{$8 \times 8$ MNIST (low vs high)}
\begin{figure}
  \begin{center}
  \includegraphics[width=0.22\textwidth]{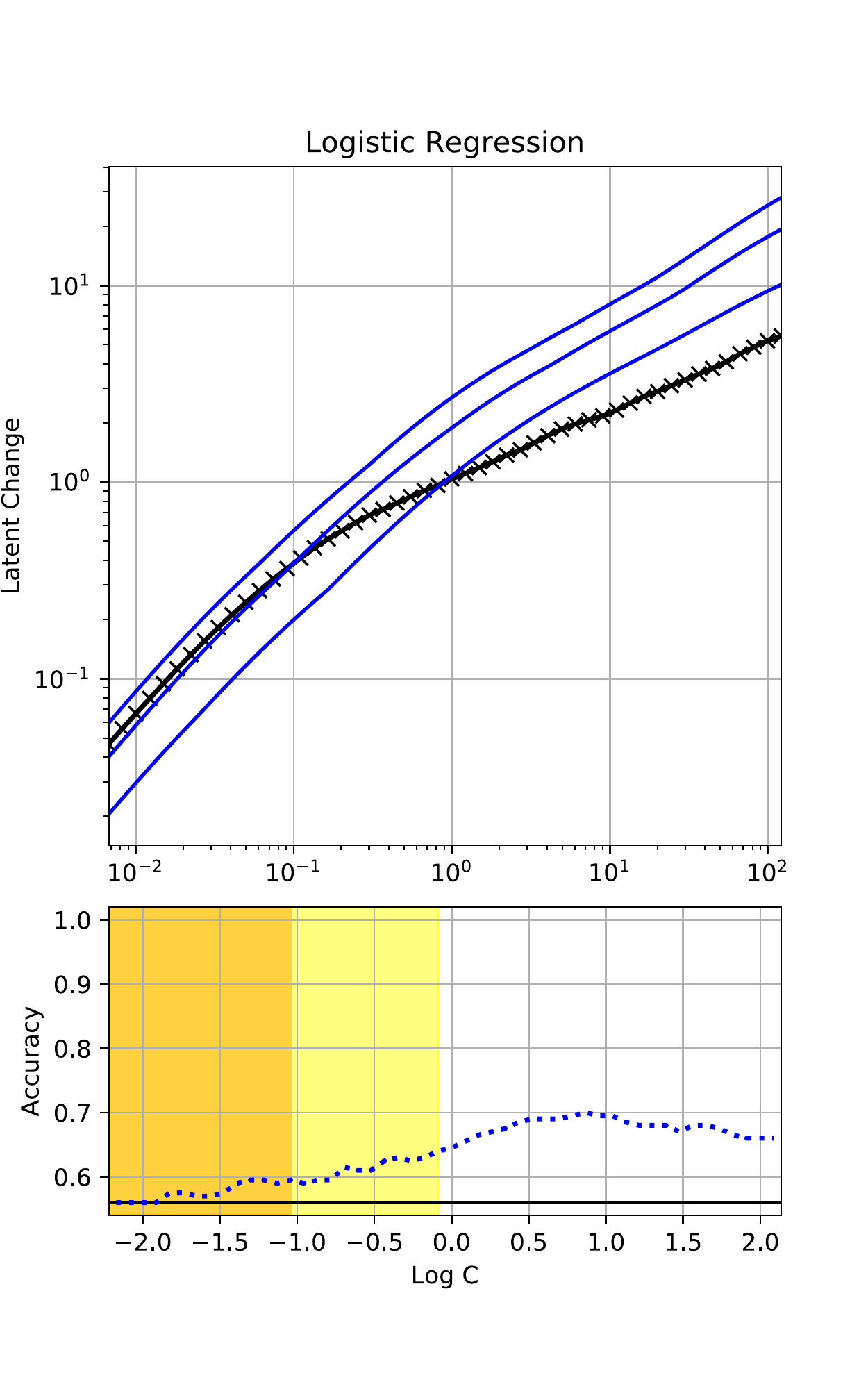}
  \includegraphics[width=0.22\textwidth]{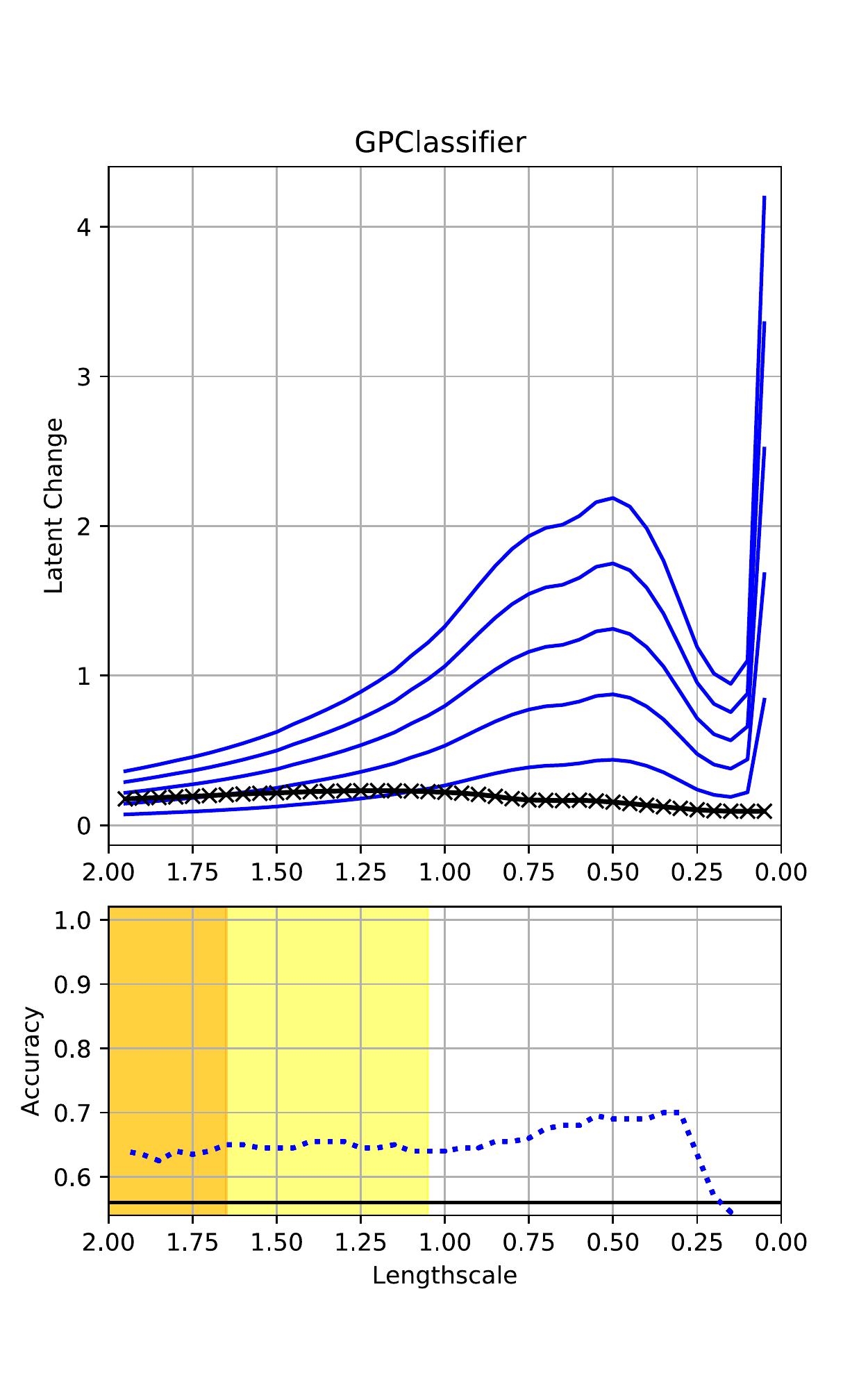}
  \end{center}
\caption{Logistic Regression (left) and GPC (right) for the MNIST low vs high digit test using just 50 training points. Upper plots: The exact maximum change in the posterior induced by changing 1, 2 or 3 input points (blue lines) for given values of the regulariser or lengthscale. The black line indicates the `confident misclassification' threshold. Lower plot: how accuracy varies wrt the regulariser or lengthscale. The yellow/orange areas indicate regions in which two/three pixels need changing to cause a confident misclassification.}
  \label{split_50data}
\end{figure}
We also ran the algorithm on the $8 \times 8$ MNIST example, using the ${0,1,2,3,4}$ or ${5,6,7,8,9}$ classification problem, using 50 training points. This is a relatively difficult problem (compared to simple digit pair comparison) especially given the low numbers of training points. Note only 43 pixels of the 64 are used by the classifier (as the remaining pixels were almost constant across training points and so excluded). The logistic regression (LR) algorithm achieves 70\% performance (see figure \ref{split_50data}). The regularisation for LR corresponds to a smaller value of $C$, while for GPC involves longer lengthscales. The GPC has similar accuracies. Of more interest is the bound on the effect one or two pixel changes can cause. Both algorithms seem to have similar results. However the regions in which there is more proven robustness (i.e. at least three pixels need changing, orange in the figure, appear to be somewhat less accurate in the LR case. Clearly the more regularised classifiers are those with greater robustness to adversarial examples, but are also potentially those with the worst accuracy.

\subsection*{Synthetic dataset}

\begin{figure}
  \begin{center}
  \includegraphics[width=0.25\textwidth]{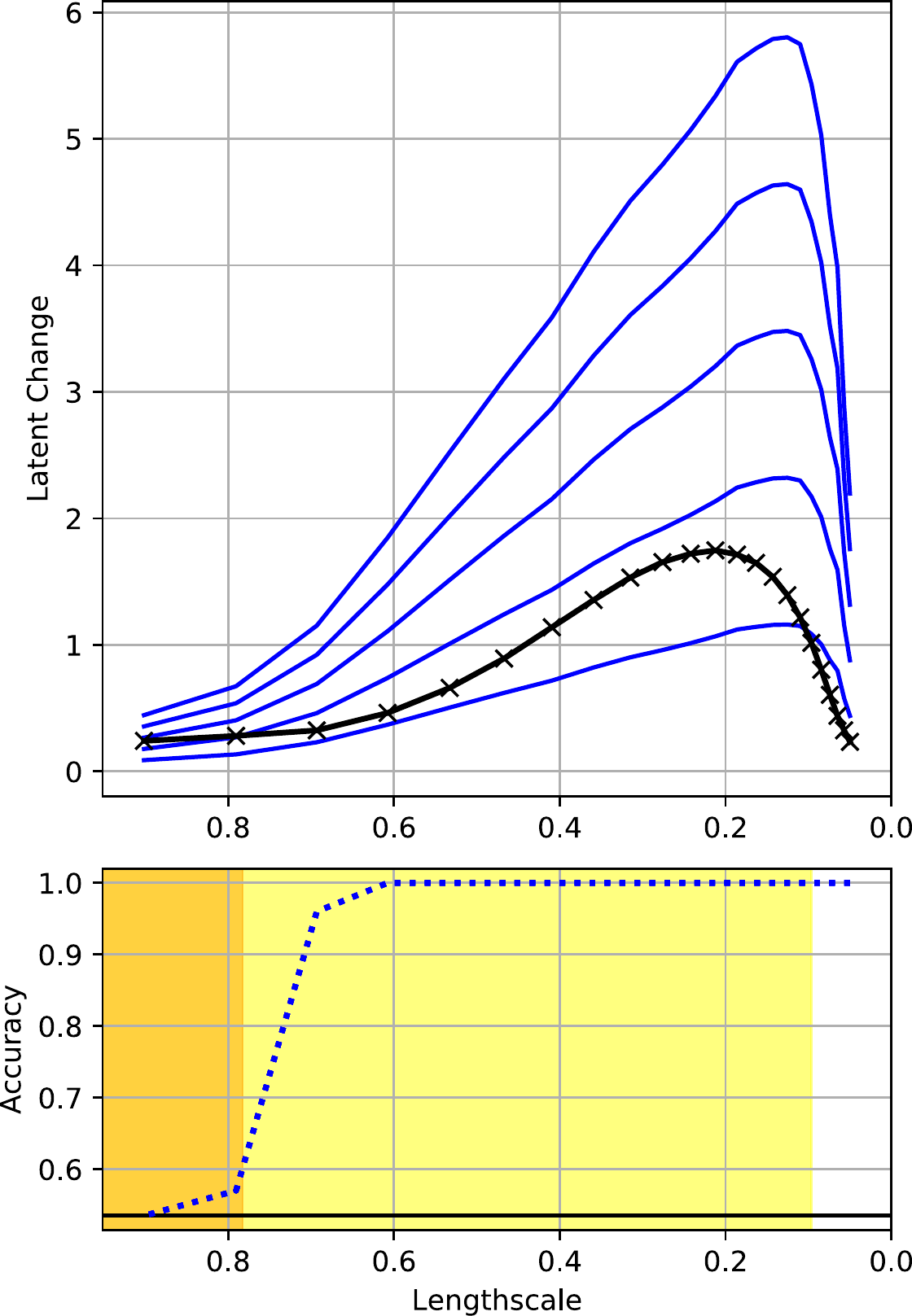}
  \end{center}
\caption{Synthetic dataset using 50 training points and the GPC. The upper plot shows upper bounds (blue lines) on the change in the posterior induced by changing 1,2,3,4 or 5 input points. The black line indicates the `confident misclassification' threshold. The lower plot shows the classifier's accuracy vs lengthscale. The yellow, orange, red and dark red areas indicate regions in which two,three,four or five pixels respectively need changing to cause a confident misclassification. Here there is a trade off between accuracy and robustness.}
  \label{synth}
\end{figure}

Figure \ref{synth} illustrates the results from the non-separable synthetic data using GPC. Note how as the bounds improve the accuracy worsens.

\subsection*{Effect of number of splits and inducing points on bound and runtime (Credit dataset)}
We also test the effect of the number of slices on the credit dataset. (400 training points, 200 test points, lengthscale of 2, 4 inducing points), we reach an accuracy of 80\%. Table \ref{effectofsplitscredit} summaries this for up to 100 slices. As for the MNIST example, more slices improve the bound but also take more computation. We also looked at the use of inducing points on this dataset (400 training points, 200 test points, lengthscale of 2, 100 slices). Table \ref{effectofinducingscredit} details these results. These also follow the pattern of the MNIST data: fewer inducing inputs is associated with a (slight) reduction in accuracy, but with improvements in the bound.

\end{document}